\documentclass[11pt, a4paper, copyright, gdm]{google}

\usepackage[authoryear, sort&compress, round]{natbib}

\usepackage{algorithmic}
\usepackage{algorithm}
\usepackage{cleveref}
\usepackage{todonotes}
\usepackage{comment}
\usepackage{wrapfig}

\usepackage{subcaption}
\usepackage{tikz}
\usetikzlibrary{positioning, arrows.meta, shapes.callouts, calc, backgrounds, shadows}

\usepackage{pgfplots}
\usepackage{pgfplotstable}
\pgfplotsset{compat=1.18}
\usepgfplotslibrary{groupplots, fillbetween}

\usepackage[many]{tcolorbox}

\definecolor{promptbg}{gray}{0.97}      %
\definecolor{promptborder}{gray}{0.75}  %

\newcommand{\pro}{\textit{Gemini-3.1-pro-preview}}
\newcommand{\flash}{\textit{Gemini-3.5-flash}}
\newcommand{\flashlite}{\textit{Gemini-3.5-flash-lite}}

\newtcolorbox{prompt}[1]{
  enhanced,
  breakable,
  colback=promptbg,
  colframe=promptborder,
  boxrule=0.6pt,
  arc=4pt,
  left=8pt, right=8pt, top=6pt, bottom=6pt,
  before=\noindent\par\smallskip,
  after=\par\smallskip,
  fonttitle=\bfseries\footnotesize\ttfamily,
  coltitle=black,
  colbacktitle=white,
  attach boxed title to top left={yshift=-2.5mm, xshift=4mm},
  boxed title style={boxrule=0.6pt, colframe=promptborder, arc=2pt},
  title={#1},
  before upper={\small\ttfamily\raggedright\parskip=0.3\baselineskip\obeylines}
}

\newcommand{\hotpotqa}{\textsc{HotpotQA}}

\newcommand{\cmark}{\ding{51}}
\newcommand{\xmark}{\ding{55}}


\uselogo{}

\title{Decentralized LLM Workflows via Sequential Auctions}

\correspondingauthor{imgemp@google.com}

\reportnumber{} %

\renewcommand{\today}{2026-10-15}

\author[1]{Ian Gemp}
\author[1]{Shubhendu Trivedi}
\author[1]{Adam Fisch}
\author[1]{Fantine Huot}
\author[1]{Michael Kaisers}
\author[1]{Mirella Lapata}
\author[1]{William W. Cohen}
\author[2]{Alekh Agarwal}
\author[1,3]{Kate Larson}
\author[1]{Jacob Eisenstein}

\affil[1]{Google DeepMind}
\affil[2]{Google Research}
\affil[3]{University of Waterloo}

\begin{abstract}
We present a scalable framework for solving multi-step tasks through a decentralized market of self-interested agents. Prior theoretical work demonstrated that agents organized through sequential auctions can converge to a globally optimal policy in a Markov decision process (MDP), but relied on the restrictive assumption of the existence of perfect clones as a model of perfect competition. In reality, exact clones drive agent profit to zero, violating the natural risk-reward incentives of participants. Conversely, we empirically demonstrate that the opposite extreme, absence of competition, also leads to market failure and value extraction collapse.
Between these two extremes, we investigate how heterogeneous agents can survive in markets that span the critical middle ground of \emph{imperfect competition}. To be profitable, agents learn to strategically shade bids to secure positive individual profit margins, while still ensuring that the marketplace functions effectively enough so that collaborative tasks posed by market consumers can still be solved. We formulate and evaluate four bidding algorithms (Mirror Descent, Thompson Sampling, Bayes-Optimal, and PPO) across three auction protocols (1st-Price, 2nd-Price, and Credit-Conserving Vickrey). Across a diverse set of domains, we demonstrate that imperfect competition successfully sustains necessary agent profit margins while maintaining high task success rates at low costs.
\end{abstract}

\begin{document}

\maketitle

\section{Introduction}
\label{sec:introduction}

Modern AI systems increasingly face tasks that require composing multiple specialized capabilities---dynamically routing queries across varying LLM sizes to optimize inference costs, gathering and synthesizing evidence from disparate sources, or dispatching subtasks to external tools. The dominant paradigm centralizes this composition in a single controller (an ``orchestrator'' LLM or a fixed computation graph), which must explicitly plan the sequence of operations and assign credit to each component~\citep{hong2024metagpt,shen2023hugginggpt}.

An alternative, inspired by economic theory, is to let a \emph{society} of specialized agents coordinate through a market mechanism. To solve temporal credit assignment across multi-step trajectories without hierarchical oversight, we draw on the foundational \emph{Bucket Brigade} principle~\citep{holland1985properties,bartholdi1996production}: each agent purchases the current task state from the preceding agent in an auction, transforms the state, and resells the updated state to the next downstream agent. Terminal task rewards earned at completion naturally propagate backward through the chain of resale prices, rewarding upstream contributors proportionally to their marginal value added. 

To ground this decentralized economy in a reinforcement learning framework, we build on the work of \citet{chang2020decentralized}. They proved that a Vickrey (second-price) auction perfectly aligns each agent's locally optimal bidding strategy with the globally optimal MDP policy. However, translating this theoretical breakthrough to real-world AI societies reveals a critical gap: their optimal equilibrium relies on the restrictive assumption of perfect competition via artificially cloned agents.

In practice, this perfect-clone assumption breaks down due to two fundamental economic realities. First is a \emph{participation barrier}: with exact clones, the expected highest and second-highest bids are identical. The winning agent pays exactly their true valuation, driving their net profit to strictly zero. Because participating in a market carries inherent risk, agents will simply sit out rather than opt in to risk their resources for a net-zero expected return. Second is the \emph{anti-cloning incentive}: perfect competition will not naturally arise because no rational actor will clone their own agent (which cannibalizes their own profit margin), and participants will actively protect their models' intellectual property from being copied by competitors.

If perfect clones are theoretically unprofitable and strategically irrational, does the market collapse? We first demonstrate that the opposite extreme---absence of competition (a \emph{monopsony} at any step in the task chain)---does indeed cause the market to fail. When an agent is the sole eligible buyer for a given state, they act as a monopsonist, bidding the absolute minimum required to win. Without competitive pressure, the clearing price collapses, and downstream value is not efficiently extracted and passed upstream. This breaks the sequential credit assignment required to sustain a multi-step pipeline, as upstream agents are starved of the revenue needed to offset their own inference costs.

To bridge this gap, we adapt the framework of \citet{chang2020decentralized} to societies of Large Language Models (LLMs), formulating a decentralized, sequential forward auction mechanism tailored to the operational realities of modern LLM pipelines. In doing so, we shift the theoretical focus to the critical, realistic middle ground: \emph{imperfect competition}. We model societies populated by LLM agents that occupy similar capability niches but are differentiated by heterogeneous inference costs, prompt engineering, and localized heuristics. Because off-the-shelf language models are notoriously miscalibrated judges of their own success probabilities and token costs~\citep{fradkin2026marketbench}, agents cannot rely on zero-shot prompting to evaluate states. Instead, each agent learns a lightweight neural valuation policy via reinforcement learning to navigate the tradeoff between winning the right to act and shading their bids to secure a positive profit margin.

In this paper, we present a comprehensive empirical study of decentralized LLM routing under imperfect competition. Our core contributions include:

\begin{enumerate}[leftmargin=*]
  \item \textbf{Decentralized LLM Markets and Imperfect Competition:} We introduce an architectural and algorithmic framework for sequential auction MDPs in LLM societies, demonstrating experimentally that imperfect competition is sufficient to sustain positive profit margins and enable viable multi-agent pipelines under traditional auction mechanisms.
  \item \textbf{Monopsony Failure Validation:} We reveal empirically through a targeted synthetic domain that absence of competition collapses the market, establishing imperfect competition as a viable operational regime under traditional auction mechanisms.
  \item \textbf{Algorithmic Bidding Strategies:} We formulate and benchmark four bidding algorithms---Mirror Descent, Thompson Sampling, Bayes-Optimal, and PPO---demonstrating how individual agents can efficiently learn optimal pricing and bid-shading policies.
  \item \textbf{Extensive Benchmarking:} We deploy our framework across diverse LLM domains (including a math reasoning task and newly-constructed synthetic news market). Through market density sweeps across 1st-Price, 2nd-Price, and Credit-Conserving Vickrey (CCV) auctions (Fig~\ref{fig:succ:rates}\&\ref{fig:translation_n4}), we show that imperfect competition is sufficient to sustain positive agent profits while simultaneously achieving higher task success rates (accuracy) than centralized orchestration baselines.
\end{enumerate}

\section{Background}
\label{sec:background}

To formalize the decentralized ecosystem just described, we model the multi-agent task pipeline as a sequential auction, in line with previous work~\citep{chang2020decentralized}. In this framework, self-interested agents bid for the right to progressively transform an initial state $s_0$ into a terminal solution $s_T$. Below, we define the underlying market MDP, the local profit incentives of the agents, and the idealized theoretical clearing rules that enable optimal credit assignment.

\subsection{The Sequential Spot Market MDP}
\label{sec:mdp_generic}

We model decentralized task routing as a finite horizon ($T$) sequential Markov Decision Process (MDP) defined by the tuple $(\mathcal{S}, \mathcal{A}, \mathcal{P}, R, \gamma)$. Rather than relying on a centralized policy $\pi(a|s)$ to select actions, the action space $\mathcal{A}$ is defined by a society of independent agents. For intuition, consider a gridworld: instead of a single controller choosing to move up or down, specialized ``Up'' and ``Down'' agents compete in an auction for the right to act on the current state $s_t$.

At each step $t$, a registry of agents evaluates $s_t$ to determine their eligibility. Eligible agents submit bids, and the highest bidder wins the right to process the state. The state transition is defined by the winning agent $i$'s localized action $a_t$, which generates the next state $s_{t+1}$ and yields an environmental reward $r(s_t, a_t)$. The process is repeated until a final state $s_T$ is reached. Social welfare is defined as the sum of discounted rewards.

Agents optimize for their individual net utility, or \emph{profit}, over the course of a transaction. For an agent winning the auction at step $t$, the profit $U_{i,t}$ is defined as the sum of the environmental reward and the net flow of capital:
\begin{equation}
  U_{i,t} = r(s_t, a_t) + P_\text{in}^{(t)} - P_\text{out}^{(t)}
  \label{eq:profit}
\end{equation}
where $P_\text{out}^{(t)}$ is the clearing price the agent pays to acquire $s_t$, and $P_\text{in}^{(t)}$ is the revenue the agent receives when they subsequently auction the transformed state $s_{t+1}$ to the next downstream agent.

\subsection{Auction Mechanisms and Credit Assignment}
\label{sec:auction_mechanisms_ideal}

The rules governing the clearing price $P_\text{out}^{(t)}$ and the subsequent revenue $P_\text{in}^{(t)}$ depend strictly on the chosen auction mechanism. \citet{chang2020decentralized} established a foundational bridge between reinforcement learning and mechanism design by demonstrating that under a Vickrey (second-price) auction, the globally optimal MDP policy constitutes a Nash equilibrium for myopic agents maximizing their local transaction profit. Their result relies on two distinct theoretical properties:
\begin{enumerate}[leftmargin=*]
    \item \textbf{Truthful Bidding via Myopia:} Assuming agents are myopic and optimize strictly for the immediate transaction, the Vickrey auction guarantees that bidding one's true valuation ($b_i = v_i$) is a weakly dominant strategy. Thus, an agent's bid reveals its true estimated action-value, $Q_i(s_t)$.
    \item \textbf{Perfect Credit Conservation via Clones:} To ensure the market clears at the optimal state-value without requiring external platform subsidies, \citet{chang2020decentralized} in their Section 5.4 build on the assumption that the market is populated with exact clones of every agent. With perfect competition via exact clones, the highest and second-highest bids are identical ($b_{(1)} = b_{(2)}$). Because the Vickrey auction charges the second-highest bid, the market clearing price exactly extracts the maximum Q-value: $P_\text{out}^{(t)} = \max_i Q_i(s_t) = V^*(s_t)$.
\end{enumerate}

When the clearing price perfectly tracks $V^*(s_t)$, the revenue passed upstream acts as an exact Bellman backup. Specifically, if the downstream auction at step $t+1$ clears at $V^*(s_{t+1})$, the upstream agent receives $P_\text{in}^{(t)} = \gamma V^*(s_{t+1})$.\footnote{\citet{chang2020decentralized} discount the payment by $\gamma$ to match standard discounted returns in RL.} By maximizing their local profit $U_{i,t}$, agents naturally align their economic incentives with the global MDP objective, recursively propagating credit back to the initial state. However, this theoretical alignment guarantee unravels if either of the above assumptions is unmet. Hence, we aim to study the robustness of a heterogeneous LLM market MDP empirically.

\section{Decentralized LLM Markets and Imperfect Competition}
\label{sec:problem_formulation}

Applying the market MDP to LLM societies requires formalizing inference costs within the reward structure and relaxing the idealized assumption of perfect competition made in previous work.

\subsection{LLM Transformations and Inference Costs}
\label{sec:llm_mdp}

In an LLM society, the state $s_t$ represents the current task context, comprising the text or structured payload (e.g., a natural language query, an intermediate translation, or a JSON object containing accumulated evidence). The action space corresponds to an agent's localized LLM invocation, transforming the payload to generate $s_{t+1}$. 

Crucially, taking this action incurs a local inference cost $c_{i,t}$ for the agent (e.g., API token costs). We formalize this by defining the environmental reward during intermediate steps as strictly negative: $r(s_t, a_t) = -c_{i,t}$. If the agent's action at step $T-1$ successfully resolves the task and transitions the environment to a terminal state $s_T$, the reward $r(s_{T-1}, a_{T-1})$ incorporates both the negative inference cost and the positive terminal reward provided by the evaluator. Because the task is complete, no subsequent auction occurs, yielding $P_\text{in}^{(T-1)} = 0$.

\subsection{The Participation Barrier and the Clone Assumption}\label{sec:participation}

We now address the problem of cloning agents in a market MDP. Under the assumption of perfect clones, the equilibrium clearing price exactly equals the winner's true valuation ($P_\text{out}^{(t)} = b_{(2)} = v_i$), hence the winner's expected utility is strictly zero ($\mathbb{E}[U_{i,t}] = 0$). An agent maximizing expected profit has no strict incentive to opt into the market, since abstention already achieves zero profit.

Furthermore, no rational agent would choose to clone themselves, as doing so strictly cannibalizes their profit margin. Conversely, participants will actively protect their models from being copied by competitors. Consequently, real LLM spot markets are bounded to operate under \emph{imperfect competition}, where $b_{(1)} > b_{(2)}$. It is precisely this structural gap that allows the winner to secure a positive profit margin ($U_{i,t} > 0$), overcoming the participation barrier.

\subsection{Defining Agent Heterogeneity}
\label{sec:agent_competition}

To operationalize the regime of \emph{imperfect competition}, we define how agents differ within the market. In our framework, all agents participate in a unified market and can theoretically bid on any state; however, they act as imperfect substitutes, organically specializing based on their localized capabilities. We parameterize this heterogeneity across two primary axes:
\begin{enumerate}[leftmargin=*]
    \item \textbf{Inference Cost ($c_{i,t}$):} Agents use different LLMs. For example, an agent backed by a $<$1B parameter LLM incurs a much lower base inference cost than one backed by a $>$100B parameter LLM, allowing it to bid competitively on lower-margin tasks.
    \item \textbf{Transition Reliability:} Agents are differentiated by distinct prompt engineering techniques, identity embeddings, and asymmetric data access (e.g., access to different document corpora). These differences cause their state transitions ($s_t \to s_{t+1}$) to vary in quality and relevance. A more capable agent is more likely to generate a state that attracts high downstream bids, thereby raising its expected future revenue.
\end{enumerate}

Because of these variances, even among agents attempting to fulfill the exact same role in a task, true valuations naturally diverge ($v_1 \neq v_2$). This structural gap allows the most efficient agent to shade its bid below its own true valuation while still outbidding the rest of the market, thereby securing a positive profit margin and incentivizing participation in the market.

\subsection{Auction Mechanisms under Imperfect Competition}
\label{sec:auction_mechanisms}

We study heterogeneous markets under the same three auctions discussed in~\citep{chang2020decentralized}:
\begin{itemize}[leftmargin=*]
    \item \textbf{2nd-Price (Vickrey):} The downstream buyer pays $b_{(2)}$, but the upstream seller receives $b_{(1)}$. The marketplace platform must act as a market maker, subsidizing the deficit ($b_{(1)} - b_{(2)}$).
    \item \textbf{Credit-Conserving Vickrey (CCV):} The buyer pays $b_{(2)}$, and the upstream seller receives $b_{(2)}$. This avoids subsidies, but fails to pass the full winning valuation upstream when $b_{(1)} > b_{(2)}$.
    \item \textbf{1st-Price (Bucket Brigade):} The winner pays their exact bid ($P_\text{out}^{(t)} = b_{(1)}^{(t)}$), and the seller receives the highest downstream bid ($P_\text{in}^{(t)} = \gamma b_{(1)}^{(t+1)}$). Credit conservation is structurally maintained without external subsidies. However, to secure a profit, agents must learn to strategically shade their bids below their true valuations.
\end{itemize}

Because standard Vickrey auctions require subsidies and CCV passes minimal value to upstream agents, we prioritize the simpler 1st-Price auction; we justify this choice in Section~\ref{sec:exp_translation}.

\subsection{Strategic Bidding in 1st-Price LLM Markets}
\label{sec:llm_bidding_formulation}

In a 1st-price auction, an agent must bid strategically to balance win probability against profit margin. To maximize expected profit, agent $i$ must first estimate its true valuation of the state, $v_i(s_t)$, which amounts to the expected future revenue minus its local inference costs for applying its transformation $a_t$. Following the definitions of $P_\text{in}^{(t)}$ and $r(s_t, a_t)$ from above, the valuation is:
\begin{equation}
v_i(s_t) = \mathbb{E}_{s_{t+1}\sim \mathcal{P}(\cdot\mid s_t, a_t)}\big[\, P_\text{in}^{(t)} + r(s_t,a_t) \,\big]
= \mathbb{E}_{s_{t+1}\sim \mathcal{P}(\cdot\mid s_t, a_t)}\big[\, \gamma\, b_{(1)}(s_{t+1}) - c_{i,t} \,\big]. \label{eq:valuation}
\end{equation}
Two agents eligible for the same state may differ in valuation through both their inference cost~$c_{i,t}$, and through the expected downstream bid $b_{(1)}(s_{t+1})$ that their different output states attract, i.e.,~the two axes of heterogeneity in Section~\ref{sec:agent_competition}.
If the agent bids its true valuation, $b_i(s_t) = v_i(s_t)$, its ex-post utility 
is zero. Consequently, a rational bidder must shade its bid below its private 
valuation to secure a positive expected surplus \citep{krishna2009auction}. We parameterize the bidding policy via a state-dependent 
multiplicative shading factor, $\xi(s_t) \in [0, 1)$, yielding the shaded bid 
$b_i(s_t) = \xi(s_t) v_i(s_t)$. The ex-post surplus upon winning is thus 
$(1 - \xi(s_t)) v_i(s_t)$, highlighting the fundamental trade-off of 1st-price 
auctions: lowering $\xi(s_t)$ widens the conditional profit margin but strictly 
decreases the probability of winning.
Despite this strategic shading, recent analysis shows that 1st-price agentic markets still achieve high welfare whenever competition is sufficiently strong~\citep{trivedi2026price}.

\section{Bidding Strategies}
\label{sec:bidding}

Crucially, our framework decouples the agent's core capability from its market strategy. The underlying LLMs responsible for the state transitions ($s_t \to s_{t+1}$) are utilized strictly for inference and remain frozen. The reinforcement learning process solely trains a lightweight MLP for each agent, which acts as the state-dependent, myopic bidding policy.
To survive in an imperfectly competitive market, an agent's MLP must solve two coupled learning problems at each step: accurately estimating the state's true valuation $v_i(s_t)$, and determining the optimal shade factor $\xi(s_t) \in (0,1)$ to maximize its expected profit margin. The agent's final bid is formulated as $b_i(s_t) = \xi(s_t) \cdot v_i(s_t)$. For Vickrey and CCV, we enforce truthful bidding, $\xi(s_t)=1$.
All agents share a common neural architecture for the valuation head, which predicts both the expected downstream value $v(s_t)$ and its standard deviation $\sigma(s_t)$. The valuation head is trained via a standard Gaussian negative log-likelihood (NLL) loss to regress toward the empirical target $P_\text{in}^{(t)} - c_{i,t}$. In what follows, we drop explicit dependence of functions on state $s_t$ but this is implied.
We evaluate four distinct approaches for computing strategic bids in 1st-price auctions:

\paragraph{Mirror Descent}

The Mirror Descent (MD) strategy frames bid-shading as an online learning problem under binary feedback (win or loss). Let $\xi^*$ represent the unobserved optimal shade threshold that matches the highest competing bid. The agent models the primal loss $\ell_t(\xi)$ as the regret of the agent's margin distance from this boundary: regret from over-bidding if the auction was won, and regret from under-bidding if lost.
Using an entropic regularizer, the proximal projection yields a closed-form multiplicative weights update for the target shade factor~\citep{beck2003mirror}:
\begin{equation}
  \xi_\text{target} =
  \begin{cases}
    \xi \cdot e^{-\eta} & \text{if won} \\
    \xi \cdot e^{+\eta} & \text{if lost}
  \end{cases}
  \label{eq:shade_md}
\end{equation}
The dynamic step size is defined as $\eta = \hat{\sigma}$ where $\hat{\sigma}$ is the estimated standard deviation of the state's valuation described above.
This updated $\xi_\text{target}$ serves as a squared-error regression target for the agent's MLP, which is trained via gradient descent.

\paragraph{Thompson Sampling}

Rather than maintaining a point estimate of the shade factor, the Thompson Sampling (TS) agent models the competitive bid threshold ($\xi^*$) probabilistically using a logit-normal posterior distribution with $\mathcal{N}(0,1)$ logit prior.
At inference time, TS samples a threshold directly from this posterior to target the minimum winning bid and maximize its realized profit margin:
\begin{equation}
  z \sim \mathcal{N}(\mu_\ell, \sigma_{\ell}^2), \quad
  \xi = \text{sigmoid}(z).
  \label{eq:thompson_shade}
\end{equation}
The threshold distribution is updated via an NLL loss conditioned on the binary auction outcome. Note that when the true threshold $\xi^* > 1$ (a losing agent), the NLL updates push $\mu_\ell \to \infty$, so the sampled shade factor converges to $\xi \to 1$ ($b_i \to v_i$) while remaining bounded below $1$.

\paragraph{Bayes-Optimal Baseline}

Instead of attempting to bid precisely at a sampled threshold, we include a Bayes-Optimal baseline (Bayes Opt) that hedges across the full uncertainty of the threshold distribution. Utilizing the same logit-normal threshold posterior as Thompson Sampling, it selects the shade factor deterministically to maximize expected profit margin under its current beliefs:
\begin{equation}
  \xi = \arg\max_{\xi' \in (0,1)} (1{-}\xi') \cdot \Phi\!\left(
    \frac{\text{logit}(\xi') - \mu_\ell}{\sigma_\ell}\right)
  \label{eq:ts_profit}
\end{equation}
where $(1{-}\xi')$ represents the proportional profit margin if the auction is won, and $\Phi(\cdot)$ is the probability of winning under the posterior model of the competitive bid threshold from \cref{eq:thompson_shade}.

\paragraph{Proximal Policy Optimization (PPO)}

Finally, we evaluate a standard deep reinforcement learning approach as in~\citep{chang2020decentralized}. The PPO policy network outputs parameters $\alpha, \beta \geq 1$ to define a unimodal Beta distribution, from which the shade factor is sampled: $\xi \sim \text{Beta}(\alpha, \beta)$.
The policy is trained to maximize the empirical profit $U_{i,t}$ using the clipped surrogate objective:
\begin{equation}
  \mathcal{L}_\text{PPO} = -\min\big( \rho_t A_t,\,
    \text{clip}(\rho_t, 1-\epsilon, 1+\epsilon) A_t \big) - \beta_\text{ent} \mathcal{H}
  \label{eq:ppo_loss}
\end{equation}
where $\rho_t$ is the probability ratio between the current and old policies, $\epsilon=0.2$, $A_t = U_{i,t} - \bar{U}_{i,\text{ema}}$ is the profit advantage over a moving average baseline ($\bar{U}_{i,\text{ema}} = 0.99\bar{U}_{i,\text{ema}} + 0.01 U_{i,t}$), and $\mathcal{H}$ is an entropy bonus to prevent premature convergence ($\beta_{ent} = 0.01$).

\paragraph{State Representation and Pre-Processing}

Because the raw environmental states in the LLM domains are non-vector text payloads, they must be transformed before an agent's trainable MLP can process them to formulate a bid. To bridge raw text and policy inputs, we employ two distinct state representation paradigms:

\begin{enumerate}[leftmargin=*]
    \item \textbf{Dense Embeddings (Sec~\ref{sec:exp_translation}):} The raw text is first embedded, e.g., utilizing Gecko~\citep{lee2024gecko}, before being passed to a trainable MLP alongside agent identity features.
    \item \textbf{Centralized Assessor (Sec~\ref{sec:exp_llm_domains}):} To aid cold-starting the market and decouple routing signals from economic bidding / pricing, a centralized \emph{assessor} LLM evaluates and compresses the task context into a discrete 0--10 relevance score for each agent. Agent MLPs operate directly on these one-hot encoded scores, effectively reducing complex routing to a tractable tabular MDP via platform-side feature engineering.
\end{enumerate}

\section{Empirical Evaluation}
\label{sec:experiments}

To empirically validate the necessity of imperfect competition and evaluate our bidding strategies, we benchmark our market across a synthetic control domain and three diverse LLM environments.

\subsection{Validating the Monopsony Failure (Synthetic Chain Translation)}
\label{sec:exp_translation}

\begin{figure*}[t]
    \centering
    \begin{subfigure}[b]{0.42\textwidth}
        \centering
        \resizebox{\textwidth}{!}{%
        \begin{tikzpicture}[
            node distance=2.6cm and 2.4cm,
            lang/.style={
              draw, circle, minimum size=1cm, font=\small\bfseries,
              fill=blue!8, inner sep=2pt,
            },
            source/.style={lang, fill=orange!20, thick},
            target/.style={lang, fill=green!15, thick},
            edge/.style={
                -{Stealth[length=5pt]}, thick, color=black!70,
            },
            cost/.style={
              fill=white, inner sep=1pt, pos=0.5,
            },
            bid/.style={
              font=\itshape, text=blue!80!black, fill=white, inner sep=1pt, pos=0.5,
            },
          ]
          \node[source]                       (fr)  {Fr};
          \node[lang, right=of fr]            (en)  {En};
          \node[lang, below=2.0cm of fr]      (sp)  {Sp};
          \node[lang, right=of en]            (ge)  {Ge};
          \node[target, above right=1.3cm and 1.9cm of ge]  (ja)  {Ja};
          \node[target, below right=1.3cm and 1.9cm of ge]  (ko)  {Ko};
          
          \draw[edge] (fr) -- node[cost, above] {0.50} node[bid, below] {$b{=}0.02{\pm}0.04$} (en);
          
          \draw[edge] (fr) -- node[cost, left, pos=0.3] {0.01} node[bid, left, pos=0.7] {$b{=}0.85{\pm}0.04$} (sp);
          
          \draw[edge] (sp) -- node[cost, below right, pos=0.3] {0.01} node[bid, below right, pos=0.7] {$b{=}0.88{\pm}0.13$} (en);
          
          \draw[edge] (en) -- node[cost, above] {0.015} node[bid, below] {$b{=}0.05{\pm}0.09$} (ge);
          
          \draw[edge] (en) to[bend left=12] node[cost, above] {0.03} node[bid, below] {$b{=}0.48{\pm}0.16$} (ja);
          
          \draw[edge] (en) to[bend right=12] node[cost, below] {0.02} node[bid, above] {$b{=}0.38{\pm}0.34$} (ko);
          
          \draw[edge] (ge) -- node[cost, above left=-1pt] {0.02} node[bid, below right=-1pt] {$b{=}0.06{\pm}0.10$} (ja);
          
          \draw[edge] (ge) -- node[cost, below left=-1pt] {0.02} node[bid, above right=-1pt] {$b{=}0.00{\pm}0.00$} (ko);
          
          \draw[dashed, red!60, thick, rounded corners, ->]
            ([xshift=-6pt, yshift=6pt] fr.north west)
            .. controls ++(0,1.0) and ++(0,1.0) ..
            node[above, font=\itshape, red!80] {Task 1 ($w{=}2$)}
            ([xshift=6pt, yshift=6pt] ja.north east);
          \draw[dashed, red!60, thick, rounded corners, ->]
            ([xshift=-6pt, yshift=-6pt] fr.south west)
            .. controls ++(0,-1.0) and ++(0,-1.0) ..
            node[below, font=\itshape, red!80] {Task 2 ($w{=}1$)}
            ([xshift=6pt, yshift=-6pt] ko.south east);
        \end{tikzpicture}%
        }
        \caption{Domain graph \& bids}
        \label{fig:translation_graph}
    \end{subfigure}
    \hfill
    \begin{subfigure}[b]{0.55\textwidth}
        \centering
        \pgfplotstableread[col sep=comma]%
  {figures/translation/274227591_274238681/success_rate_nc1.csv}\srNCone
\pgfplotstableread[col sep=comma]%
  {figures/translation/274227591_274238681/success_rate_nc2.csv}\srNCtwo

\definecolor{srBayesC}{RGB}{230,159,0}
\definecolor{srCCVC}{RGB}{110,110,110}
\definecolor{srMDC}{RGB}{0,114,178}
\definecolor{srPPOC}{RGB}{213,47,47}
\definecolor{srTSC}{RGB}{0,150,64}
\definecolor{srVickC}{RGB}{25,25,25}

\tikzset{
  srBayes/.style={srBayesC, line width=0.7pt},
  srCCV/.style={srCCVC, line width=0.7pt, densely dashed},
  srMD/.style={srMDC, line width=0.7pt},
  srPPO/.style={srPPOC, line width=0.7pt},
  srTS/.style={srTSC, line width=0.7pt},
  srVick/.style={srVickC, line width=0.7pt},
}

\begin{tikzpicture}

\def\srband#1#2#3#4{%
  \addplot[draw=none, forget plot, name path=#3#2lo]
    table[x=episode, y=#2_lo] {#1};
  \addplot[draw=none, forget plot, name path=#3#2hi]
    table[x=episode, y=#2_hi] {#1};
  \addplot[#4, opacity=0.15, forget plot] fill between[of=#3#2lo and #3#2hi];
}
\def\srmean#1#2#3{\addplot[#3] table[x=episode, y=#2_mean] {#1};}

\begin{groupplot}[
  group style={
    group size=2 by 1,
    horizontal sep=0.28cm,
    ylabels at=edge left,
    yticklabels at=edge left,
  },
  scale only axis=true,
  width=0.385\linewidth,
  height=0.27\linewidth,
  xlabel={Episode},
  ylabel={Success rate},
  xmin=0, xmax=5000,
  ymin=-0.03, ymax=1.01,
  xtick={0,2000,4000},
  xticklabels={0,2k,4k},
  ytick={0,0.5,1},
  yticklabels={0,.5,1},
  grid=major,
  grid style={gray!25, line width=0.25pt},
  tick label style={font=\tiny},
  label style={font=\tiny},
  xlabel style={yshift=2pt},
  ylabel style={yshift=-3pt},
  title style={font=\tiny, yshift=-4pt},
  axis line style={gray!60, line width=0.3pt},
  tick style={gray!60},
  every axis plot post/.append style={line join=round},
  legend cell align=left,
  legend columns=-1,
  legend to name=srSuccessLegend,
  legend image code/.code={\draw[#1] plot coordinates {(0cm,0cm) (0.115cm,0cm)};},
  legend style={
    draw=none, fill=none,
    font=\fontsize{5}{5.4}\selectfont,
    inner xsep=0pt, inner ysep=0pt,
    /tikz/every even column/.append style={column sep=1pt},
    /tikz/every odd column/.append style={column sep=2.2pt},
  },
]

\nextgroupplot[title={$1$ competitor (monopsony)}]
\srband{\srNCone}{bayesopt}{a}{srBayesC}
\srband{\srNCone}{ccvickrey}{a}{srCCVC}
\srband{\srNCone}{mirror}{a}{srMDC}
\srband{\srNCone}{ppo}{a}{srPPOC}
\srband{\srNCone}{ts}{a}{srTSC}
\srband{\srNCone}{vickrey}{a}{srVickC}

\srmean{\srNCone}{bayesopt}{srBayes}
\srmean{\srNCone}{ccvickrey}{srCCV}
\srmean{\srNCone}{mirror}{srMD}
\srmean{\srNCone}{ppo}{srPPO}
\srmean{\srNCone}{ts}{srTS}
\srmean{\srNCone}{vickrey}{srVick}

\nextgroupplot[title={$2$ competitors}]
\srband{\srNCtwo}{bayesopt}{b}{srBayesC}
\srband{\srNCtwo}{ccvickrey}{b}{srCCVC}
\srband{\srNCtwo}{mirror}{b}{srMDC}
\srband{\srNCtwo}{ppo}{b}{srPPOC}
\srband{\srNCtwo}{ts}{b}{srTSC}
\srband{\srNCtwo}{vickrey}{b}{srVickC}

\srmean{\srNCtwo}{bayesopt}{srBayes}   \addlegendentry{Bayes Opt}
\srmean{\srNCtwo}{ccvickrey}{srCCV}    \addlegendentry{CC-Vickrey}
\srmean{\srNCtwo}{mirror}{srMD}        \addlegendentry{Mirror Descent}
\srmean{\srNCtwo}{ppo}{srPPO}          \addlegendentry{PPO}
\srmean{\srNCtwo}{ts}{srTS}            \addlegendentry{TS}
\srmean{\srNCtwo}{vickrey}{srVick}     \addlegendentry{Vickrey}

\end{groupplot}

\path (group c1r1.north west) -- (group c2r1.north east)
      coordinate[midway] (srlegendmid);
\node[anchor=south, inner sep=0pt] at ([yshift=13pt] srlegendmid)
  {\hypersetup{hidelinks}\ref{srSuccessLegend}};

\end{tikzpicture}
        \caption{Success rate over training}
        \label{fig:succ:rates}
    \end{subfigure}
    \caption{Translation Domain Validation:
    (\subref{fig:translation_graph}) The task graph showing inference costs (black) and learned Vickrey bids (blue, italic; mean $\pm$ std dev). Downstream bids reflect task sampling probabilities (e.g., \texttt{En}$\to$\texttt{Ja} expects $0.48$ because Task 1 is sampled $2/3$ of the time). The upstream \texttt{Sp}$\to$\texttt{En} bid ($0.88$) accurately approximates the downstream expected value minus local costs ($0.48 + 0.38 - 0.01 = 0.85$). (\subref{fig:succ:rates}) Success rate increases and stabilizes across all auctions under increasing market density. Competitors are differentiated by cost $c = c_b (1 + t)$ with $t \sim U(-0.5, 0.5)$ and base costs $c_b$ listed in the graph. Curves are means over $3$ independent trials and shaded regions span $\pm 1$ standard deviation; both are computed on a trailing $101$-episode rolling average.}
    \label{fig:translation_full}
\end{figure*}
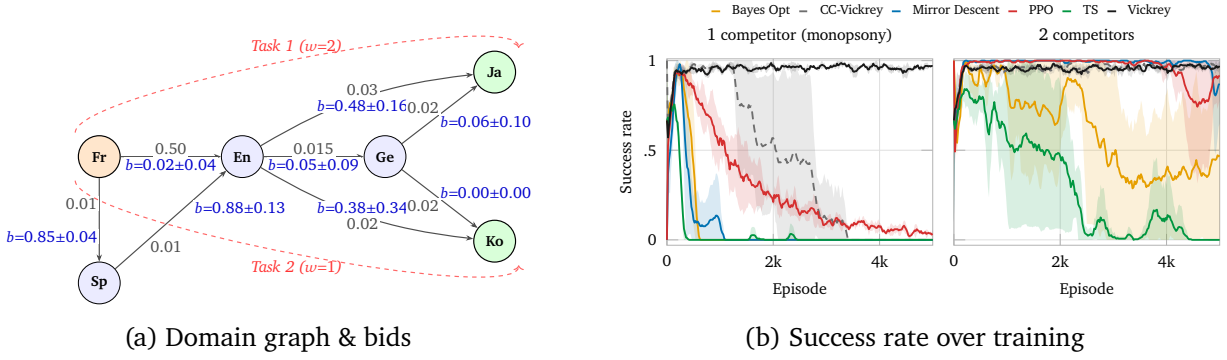

We first evaluate the framework on a synthetic multi-step translation domain. In this environment, states are constructed by concatenating two $64$-dimensional embeddings, one identifying the original task and the other the current language. Agents act as bilingual translators stationed on the edges of a graph (Figure~\ref{fig:translation_graph}), possessing distinct inference costs $c_i$. The market must route payloads from a source node to target nodes with terminal reward $\$1$ by discovering the most cost-efficient topological sequence (e.g., bypassing the expensive $\$0.50$ direct \texttt{Fr}$\to$\texttt{En} edge in favor of the $\$0.02$ \texttt{Fr}$\to$\texttt{Sp}$\to$\texttt{En} path). We evaluate all three proposed auctions: 1st-price, 2nd-price, and CCV.

This domain allows us to test the extremes of market competition. We configure the environment with a single agent per edge ($N=1$). In this regime, intermediate nodes frequently act as \emph{monopsonies}. This occurs either structurally (e.g., the \texttt{Sp} has only one outgoing edge) or task-conditionally (e.g., \texttt{Ge} has two edges, but only one is viable for a given task, rendering the competitor's bid zero).

The presence of an uncontested buyer exposes the failure modes of each auction type. In a standard 2nd-price (Vickrey) auction, the monopsonist bids truthfully but pays the floor ($0$); the platform must inject unacceptable subsidies (e.g., $\$0.88$ at \texttt{Sp}) to pay the upstream seller the first price. Under CCV, the buyer pays $0$ and the upstream seller receives $0$. Under a 1st-price auction, the monopsonist actively shades its bid down to the floor, passing a vanishingly small value upstream. 

For CCV and 1st-price auctions, this failure to pass value backward triggers a cascading network collapse. Upstream agents (e.g., \texttt{Fr}$\to$\texttt{Sp} and \texttt{En}$\to$\texttt{Ge}) receive $P_\text{in} \approx 0$, fail to offset their inference costs, and go bankrupt. Their exit removes competition at earlier nodes (e.g., the \texttt{En} node loses the \texttt{En}$\to$\texttt{Ge} buyer), turning those preceding nodes into new task-conditioned monopsonies that also drop their bids to zero. Because every edge incurs a positive inference cost, this backward unraveling inevitably starves every branch of the network, guaranteeing total market failure.

When we transition to an \emph{imperfect competition} regime by populating each edge with $N>1$ agents possessing slight cost variances, the monopsonies are broken. Competitive pressure forces multiple buyers to raise their bids above the floor. Value successfully flows backward through the graph, enabling upstream agents to sustain positive profit margins and restoring the optimal $\$0.04$ \texttt{Fr}$\to$\texttt{Sp}$\to$\texttt{En}$\to$\texttt{Ko} route without requiring subsidies. We discuss the poor performance of TS and Bayes-Optimal baselines with $N{=}2$ in Appendix~\ref{app:alg}. All methods succeed with $N{=}4$ (Appendix~\ref{app:add_exp}).

Crucially, this experiment validates our auction mechanics under imperfect competition. Because standard 2nd-price (Vickrey) auctions require theoretically unbounded platform subsidies, and CCV passes insufficient value to upstream agents when competition is imperfect (often starving them of the revenue needed to cover inference costs), we prioritize the simpler 1st-Price auction---where agents actively learn bid shading---as the default mechanism for subsequent experiments.

\subsection{LLM Agent Economies}
\label{sec:exp_llm_domains}

Having validated the economic mechanics, we scale the framework to diverse LLM domains: News Market (information synthesis) and Omni-MATH (mathematical reasoning).
We also examine HotpotQA in Appendix~\ref{app:add_exp}.
These domains test the market's ability to orchestrate multi-step cognitive workflows. To model imperfect competition, agents are differentiated along three axes: dynamic identity embeddings, asymmetric data access, and distinct system prompts (full prompts are detailed in Appendix~\ref{app:agent-prompts}). Unless otherwise noted, all agents share a frozen \flashlite{} backbone. Inference cost pricing is calculated using real input/output token costs.

\paragraph{Asymmetric Data Markets (News Market):} 
We construct the News Market (Appendix~\ref{app:news_market_data}) to simulate a pay-as-you-go research pipeline. Competition is driven by both asymmetric information and query strategy. The market contains four publisher categories (wikipedia, medical journal, social media, news) operating over strictly disjoint document corpora, alongside a \emph{sales agent} that synthesizes the final answer. To establish imperfect competition within each category, we deploy two competing \emph{publisher agents} per category. These intra-category competitors share the same data access but are differentiated by distinct search prompts: a \emph{precision} strategy generating narrow, entity-specific queries versus a \emph{coverage} strategy casting a broader, gap-filling net.

\begin{figure*}[t]
    \centering
    \pgfplotstableread[col sep=comma]%
  {figures/news_market/276269706/agg_success.csv}\nwSuccess
\pgfplotstableread[col sep=comma]%
  {figures/news_market/276269706/agg_inference_cost.csv}\nwCost
\pgfplotstableread[col sep=comma]%
  {figures/news_market/276269706/agg_profit.csv}\nwProfit
\pgfplotstableread[col sep=comma]%
  {figures/news_market/276269706/agg_agent_profit.csv}\nwAgent

\definecolor{nwMDC}{RGB}{0,114,178}
\definecolor{nwPPOC}{RGB}{213,47,47}

\definecolor{nwNewsC}{RGB}{204,121,167}   %
\definecolor{nwMedC}{RGB}{86,180,233}     %
\definecolor{nwSalesC}{RGB}{230,159,0}    %
\definecolor{nwSocialC}{RGB}{0,158,115}   %
\definecolor{nwWikiC}{RGB}{0,0,0}         %

\tikzset{
  nwMD/.style={nwMDC, line width=0.8pt},
  nwPPO/.style={nwPPOC, line width=0.8pt, dashed},
  nwA0/.style={line width=0.7pt},
  nwA1/.style={line width=0.7pt, dashed},
}

\begin{tikzpicture}

\def\nwband#1#2#3#4{%
  \addplot[draw=none, forget plot, name path=#3#2lo]
    table[x=episode, y=#2_lo] {#1};
  \addplot[draw=none, forget plot, name path=#3#2hi]
    table[x=episode, y=#2_hi] {#1};
  \addplot[#4, opacity=0.18, forget plot] fill between[of=#3#2lo and #3#2hi];
}
\def\nwmean#1#2#3{\addplot[#3] table[x=episode, y=#2_mean] {#1};}

\begin{groupplot}[
  group style={
    group size=4 by 1,
    horizontal sep=0.8cm,  %
  },
  scale only axis=true,
  width=0.19\linewidth,
  height=0.14\linewidth,
  xlabel={Episode},
  xmin=0, xmax=5000,
  xtick={0,2500,5000},
  xticklabels={0,2.5k,5k},
  minor xtick={1250,3750},
  grid=both,
  grid style={gray!25, line width=0.3pt},
  minor grid style={gray!25, line width=0.3pt},
  minor tick length=0pt,
  tick label style={font=\scriptsize},
  label style={font=\scriptsize},
  xlabel style={yshift=6pt},
  title style={font=\scriptsize, yshift=-7pt, text height=1.7ex, text depth=0.3ex},  %
  axis line style={gray!60},
  tick style={gray!60},
  every axis plot post/.append style={line join=round},
  legend cell align=left,
  legend columns=1,
  legend image code/.code={\draw[#1] plot coordinates {(0cm,0cm) (0.26cm,0cm)};},
  legend style={
    font=\fontsize{5.5pt}{6pt}\selectfont,
    draw=gray!55, line width=0.3pt, fill=white,
    inner xsep=1.5pt, inner ysep=0.5pt,
    row sep=-3.2pt,
    /tikz/every even column/.append style={column sep=1pt},
  },
]

\nextgroupplot[
  title={Success rate},
  ymin=-0.02, ymax=1.01,
  ytick={0,0.5,1},
  yticklabels={0,.5,1},
]
\nwband{\nwSuccess}{mirror}{a}{nwMDC}
\nwband{\nwSuccess}{ppo}{a}{nwPPOC}
\nwmean{\nwSuccess}{mirror}{nwMD}
\nwmean{\nwSuccess}{ppo}{nwPPO}

\nextgroupplot[
  title={Inference cost ($\times 10^{-3}$)},
  ymin=0, ymax=1.0,
  ytick={0,0.5,1},
  yticklabels={0,.5,1},
  y filter/.expression={1000*y},
]
\nwband{\nwCost}{mirror}{b}{nwMDC}
\nwband{\nwCost}{ppo}{b}{nwPPOC}
\nwmean{\nwCost}{mirror}{nwMD}
\nwmean{\nwCost}{ppo}{nwPPO}

\nextgroupplot[
  title={Platform profit},
  ymin=0,
  scaled y ticks=false,
  yticklabel style={/pgf/number format/fixed, /pgf/number format/1000 sep={}},
  legend style={at={(0.03,0.97)}, anchor=north west},
]
\nwband{\nwProfit}{mirror}{c}{nwMDC}
\nwband{\nwProfit}{ppo}{c}{nwPPOC}
\nwmean{\nwProfit}{mirror}{nwMD}   \addlegendentry{Mirror Descent}
\nwmean{\nwProfit}{ppo}{nwPPO}     \addlegendentry{PPO}

\nextgroupplot[
  title={Per-agent profit (MD)},
  ymin=0,
  scaled y ticks=false,
  yticklabel style={/pgf/number format/fixed, /pgf/number format/1000 sep={}},
  legend style={at={(0.02,0.985)}, anchor=north west},
]
\nwband{\nwAgent}{news0}{d}{nwNewsC}
\nwband{\nwAgent}{news1}{d}{nwNewsC}
\nwband{\nwAgent}{med0}{d}{nwMedC}
\nwband{\nwAgent}{med1}{d}{nwMedC}
\nwband{\nwAgent}{sales0}{d}{nwSalesC}
\nwband{\nwAgent}{sales1}{d}{nwSalesC}
\nwband{\nwAgent}{social0}{d}{nwSocialC}
\nwband{\nwAgent}{social1}{d}{nwSocialC}
\nwband{\nwAgent}{wiki0}{d}{nwWikiC}
\nwband{\nwAgent}{wiki1}{d}{nwWikiC}
\nwmean{\nwAgent}{news0}{nwA0, nwNewsC, forget plot}
\nwmean{\nwAgent}{news1}{nwA1, nwNewsC, forget plot}
\nwmean{\nwAgent}{med0}{nwA0, nwMedC, forget plot}
\nwmean{\nwAgent}{med1}{nwA1, nwMedC, forget plot}
\nwmean{\nwAgent}{sales0}{nwA0, nwSalesC, forget plot}
\nwmean{\nwAgent}{sales1}{nwA1, nwSalesC, forget plot}
\nwmean{\nwAgent}{social0}{nwA0, nwSocialC, forget plot}
\nwmean{\nwAgent}{social1}{nwA1, nwSocialC, forget plot}
\nwmean{\nwAgent}{wiki0}{nwA0, nwWikiC, forget plot}
\nwmean{\nwAgent}{wiki1}{nwA1, nwWikiC, forget plot}
\addlegendimage{nwA0, nwNewsC}   \addlegendentry{news}
\addlegendimage{nwA0, nwMedC}    \addlegendentry{med}
\addlegendimage{nwA0, nwSalesC}  \addlegendentry{sales}
\addlegendimage{nwA0, nwSocialC} \addlegendentry{social}
\addlegendimage{nwA0, nwWikiC}   \addlegendentry{wiki}

\end{groupplot}
\end{tikzpicture}
    \caption{Economic Viability on News Market: The decentralized market successfully sustains social welfare while delivering strictly positive cumulative profits to both individual agents and the platform. The per-agent cumulative profit is reported for Mirror Descent. Means and standard deviations are computed over $5$ trials.}
    \label{fig:news_market}
\end{figure*}
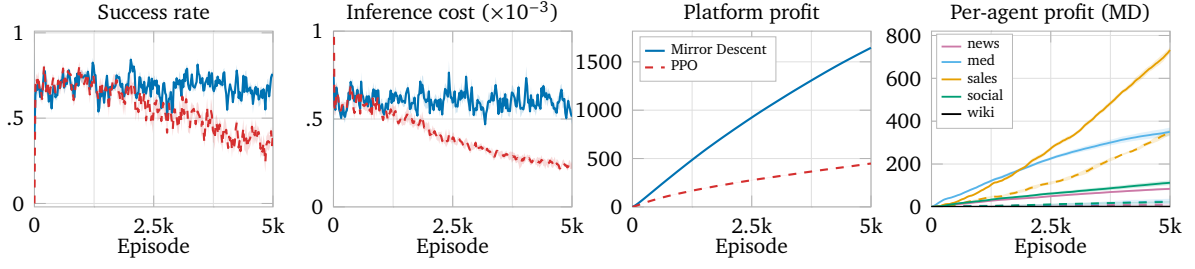

The market successfully learns to route queries to relevant publishers based on expected downstream valuation, achieving a $73.4 \pm 1.7\%$ success rate over the final $100$ episodes (\cref{fig:news_market}). Importantly, active market participants achieve positive profit margins as well, overcoming the participation barrier raised in~\Cref{sec:participation}. In comparison, a centralized ReAct baseline \citep{yao2023react} limited to the same $8$-step horizon achieves $48.4 \pm 2.8\%$ with \flashlite{} at $0.035 \pm 0.0004 \, \text{\textcent}$/task and $78.8 \pm 1.3\%$ with \pro{} at $1.060 \pm 0.368 \, \text{\textcent}$/task ($1$\textcent${=}\$0.01$). Operating exclusively with \flashlite{} agents, the market recovers $82\%$ of \pro's accuracy lift over \flashlite{} ReAct at an agent inference cost of just $0.054 \pm 0.003 \, \text{\textcent}$/task ($0.102\text{\textcent}$ including the market assessor)\textemdash a $19.7\times$ cost reduction relative to \pro{} ($10.4\times$ total). Although the market takes slightly fewer steps per task than \flashlite{} ReAct ($3.12$ vs.\ $3.23$), its higher retrieval success accumulates richer multi-document contexts for the sales agent to synthesize, trading a minor $0.067 \, \text{\textcent}$/task increase for a $+25.0\%$ accuracy gain.

\emph{Bidding Strategy Performance:} We found Mirror Descent (MD) to outperform PPO empirically, particularly in the later stages of training. MD decouples the learning of a) valuations and b) shading, providing a clear signal to each: a) regressing valuations toward realized payoffs upon winning, and b) adjusting the shade factor up or down strictly from the binary loss/win outcome. In contrast, while PPO tracks MD early in training (when shades are initialized near $0.5$ and winning agents accumulate a positive profit baseline $\bar{U}_{i,\text{ema}} > 0$), its performance degrades later as competing bids rise. When stiffer competition requires an agent to bid a higher shade $\xi_t$, exploring an increased shade that still loses ($U_{i,t} = 0$) yields a negative advantage ($A_{i,t} = -\bar{U}_{i,\text{ema}} < 0$), producing a negative policy gradient that pushes the bid \emph{down} rather than up; moreover, even when a higher shade wins, the compressed late-stage profit margin ($0 < U_{i,t} < \bar{U}_{i,\text{ema}}$) also registers as a negative advantage relative to earlier margins, causing PPO to systematically under-bid as the market matures.

\begin{table}[t]
\centering
\caption{News Market performance and inference cost per task ($1$\textcent${=}\$0.01$) including the assessor. All market participants use \flashlite. Best market mechanism is shown in \textbf{bold}.}
\label{tab:news_market_summary}
\small
\begin{tabular}{lccc}
\toprule
\textbf{Method} & \textbf{Success Rate (\%)} & \textbf{Steps / Task} & \textbf{Cost / Task (\textcent)} \\
\midrule
ReACT (\flashlite) & $48.40 \pm 2.79$ & $3.23 \pm 0.04$ & $0.0348 \pm 0.0004$ \\
ReACT (\pro) & $78.80 \pm 1.30$ & $4.37 \pm 0.12$ & $1.0599 \pm 0.3679$ \\
\midrule
Market (PPO, 1st-Price) & $40.00 \pm 2.83$ & $1.85 \pm 0.12$ & $0.0525 \pm 0.0036$ \\
\textbf{Market (Mirror Descent, 1st-Price)} & $\mathbf{73.40 \pm 1.67}$ & $\mathbf{3.12 \pm 0.11}$ & $\mathbf{0.1020 \pm 0.0044}$ \\
\bottomrule
\end{tabular}
\end{table}

\begin{table}[t]
\centering\small%
\caption{Test set performance on Omni-MATH with PPO. Sequential auctions achieve the highest overall accuracy, outperforming a prompted router at comparable cost (measured in number of refinements, which is monotonic in token use).}
\label{tab:math-results}
\begin{tabular}{lll}
\toprule
\textbf{Method} & \textbf{Accuracy} (se) & \textbf{Cost} \\
\midrule
one-shot & $69.17~{(\pm 1.44)}$ & $0$ \\
fixed $n = 1$ & $70.54~(\pm 1.42)$ & $1$ \\
fixed $n = 2$ & $71.80~(\pm 1.41)$ & $2$ \\
prompted & $72.59~(\pm 1.39)$ & $0.24$ \\
auction & $74.83~(\pm 1.36)$ & $0.28$ \\
\bottomrule
\end{tabular}
\end{table}

\textbf{Budgeting Reasoning Effort (Omni-MATH).} Our second domain tests whether the market can allocate reasoning effort adaptively across mathematical problems. Following the \textbf{g}enerator/\textbf{v}erifier/\textbf{r}efiner decomposition of MALT~\citep{motwani2024malt}, we instantiate three agents of each role and a single submitter. Generators first bid to produce a candidate solution. Verifiers then bid against one another and the submitter to assess it; if the submitter wins, the rollout terminates. Otherwise, refiners bid against one another and the submitter using the candidate and verification feedback; a winning refiner replaces the candidate and returns control to the verification stage. The submitter therefore acts as an endogenous stopping threshold: another reasoning step is purchased only when an agent values further work more highly than submitting the current answer. 

All agents use prompted \flash, with three prompts per role (\Cref{app:prompts-math}). A small multi-layer perceptron (MLP) estimates agent values from structural features and the output of a prompted \emph{assessor} that scores the current candidates. We train the value estimator and evaluate on held-out Omni-MATH problems~\citep{gao2025omnimath}, using the domains and splits from~\citet{fisch2026pandora}. As reported in Table~\ref{tab:math-results}, the auction reaches $74.83\%$ accuracy with only $0.28$
refinements per problem, outperforming fixed one- and two-refinement
baselines and a prompted adaptive orchestrator inspired by MALT, while incurring comparable cost. All comparisons with the auction-based method are statistically significant at $p<.05$ by McNemar's test.

\section{Related Work}
\label{sec:related_work}

\paragraph{Top-Down Orchestration and Routing.}
Prior work scales LLM reasoning via external scaffolding, including static graph/configuration optimizers like Maestro~\citep{wang2025maestro}, centralized cost-cascading routers like FrugalGPT~\citep{chen2024frugalgpt}, and learned small-agent routing~\citep{alazraki2026scalingsmallagentsstrategy}. Unlike our decentralized market, these systems presuppose a fixed topology or a global planner with full information, ignoring private operational costs.

\paragraph{Decentralized Agent Markets.}
The idea of building multi-step trajectories from sequential auctions was introduced by \citet{chang2020decentralized}, but their Vickrey-based framework required an awkward ``cloning" construction to preserve credit conservation. We show that 1st-price auctions can be effective, and we adapt their framework to contemporary LLM-based applications such as Omni-Math. This adaptation is what necessitates the novel learning architecture described in sections~\ref{sec:problem_formulation} and \ref{sec:bidding}.
More recently, \citet{tomasev2025virtual} propose (but do not implement) coordination of contemporary LLMs through market mechanisms, and specific market-like mechanisms are presented by \citet{qi2026economy} and \citet{liu2026markets}. Our approach is distinguished by pushing the idea of decentralization further. In particular, \citet{qi2026economy} maintain a population of agents that is selected by each agent's economic balance (agents with zero balance drop out), but the agents are not economic \emph{actors}: they do not structure their participation to optimize a local economic incentive. Concurrent work by \citet{liu2026markets} orchestrates LLMs using hierarchical reverse procurement auctions where contractors subcontract tasks. In contrast, our sequential spot market (\cref{sec:background}) resolves multi-step credit assignment through peer-to-peer state resales without top-down subcontracting trees. Another difference is that \citet{liu2026markets} centralize value estimation in a single model (the participants estimate only their own costs), while we require each participant to estimate its own expected return. This aligns with our vision of a market composed of an evolving set of niche service providers.

\paragraph{Economic Alignment Benchmarks.}
Several benchmarks evaluate whether prompt-driven LLMs act rationally in simulated retail or auction games, including Magentic Market~\citep{bansal2025magentic}, VendingBench~\citep{backlund2025vending}, CoffeeBench~\citep{sugiura2026coffeebench}, MarketBench~\citep{fradkin2026marketbench}, and Agent Bazaar~\citep{karten2026agent}. Whereas those works study whether raw LLMs natively behave like \emph{homo economicus}, we use mechanism design and RL-trained bidding policies as an algorithmic organizing principle to coordinate multi-agent pipelines.

\section{Discussion and Conclusion}
\label{sec:conclusion}

We present a decentralized market framework for LLM orchestration. Whereas prior frameworks assume perfect clones---ignoring participation incentives---we establish \emph{imperfect competition} as a viable regime. Under 1st-price auctions, agents pairing frozen LLMs with trainable bidding networks learn strategic bid-shading. This secures positive margins and conserves credit across pipelines, avoiding the unbounded subsidies of 2nd-price and the severe credit attenuation of CCV.
Empirically, complex cognitive workflows emerge organically via price discovery, bypassing top-down orchestration. Furthermore, our density sweeps confirm that scaling competition efficiently compresses individual margins while systematically increasing social welfare and task accuracy.

Our framework has several limitations: it requires sufficient competition (collapsing under monopsonies), assumes strictly sequential task transformations (precluding parallel execution), relies on myopic agents rather than globally strategic planners, and utilizes a centralized assessor rather than fully decentralized feature discovery. Future work may jointly fine-tune LLM backbones directly on market profit signals and investigate robustness against malicious bidding and collusion.
Ultimately, however, this work shows that decentralized market-based routing offers a scalable blueprint for transitioning from rigid AI controllers to open, self-sustaining agentic economies.

\bibliography{main}
\bibliographystyle{abbrvnat}

\newpage
\appendix

\section{Additional Experiments}\label{app:add_exp}

\begin{figure}[h!]
    \centering
    \includegraphics[width=0.6\linewidth]{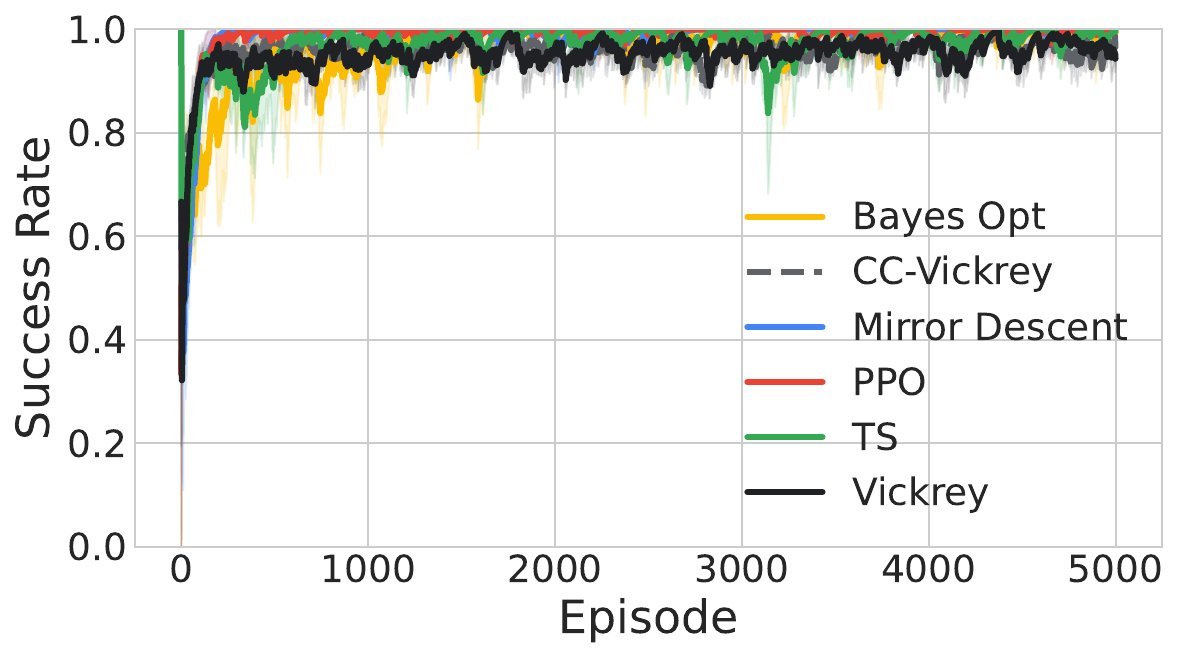}
    \caption{Translation Domain Scaling ($N=4$). When the market density is increased to 4 competitors per edge, competitive pressure is sufficiently high that even the Thompson Sampling (TS) and Bayes-Optimal baselines are able to successfully discover profitable shading margins and sustain the market. This contrasts with the $N=2$ regime shown in the main text (Figure~\ref{fig:succ:rates}, right panel), where only PPO and Mirror Descent exhibited the robustness required to stabilize the supply chain.}
    \label{fig:translation_n4}
\end{figure}

\paragraph{Translation Domain} Figure~\ref{fig:translation_n4} demonstrates that all methods are able to reach high success rates under sufficient competition.

\begin{figure*}[t]
    \centering
    \includegraphics[width=0.32\linewidth]{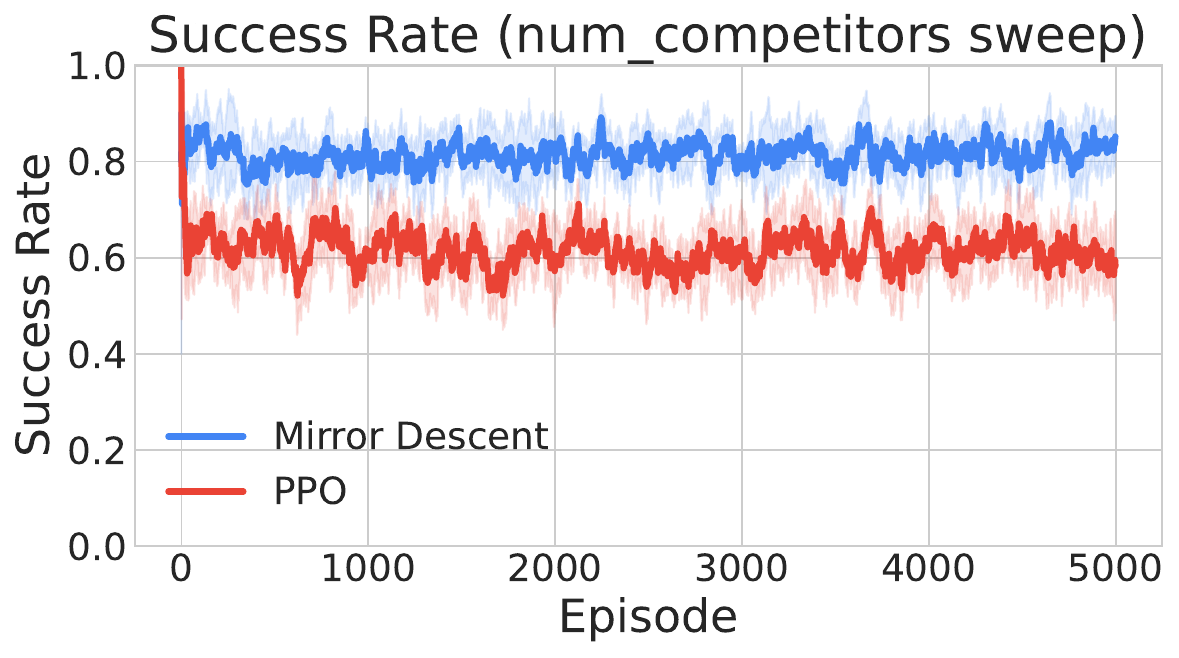}
    \includegraphics[width=0.32\linewidth]{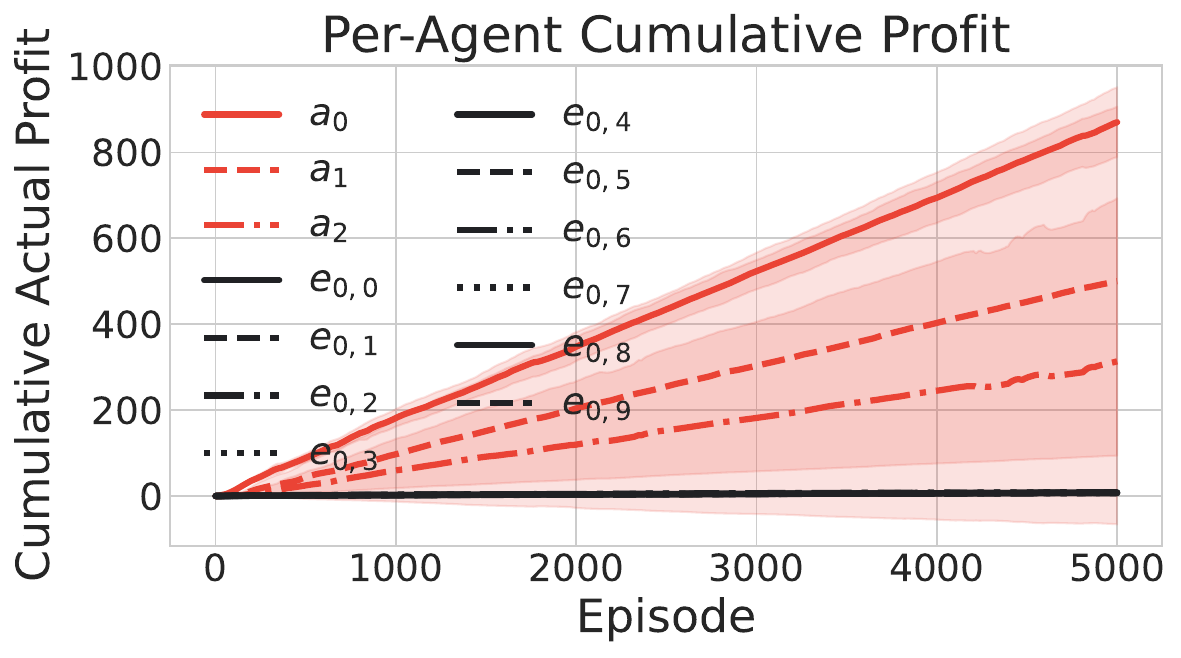}
    \includegraphics[width=0.32\linewidth]{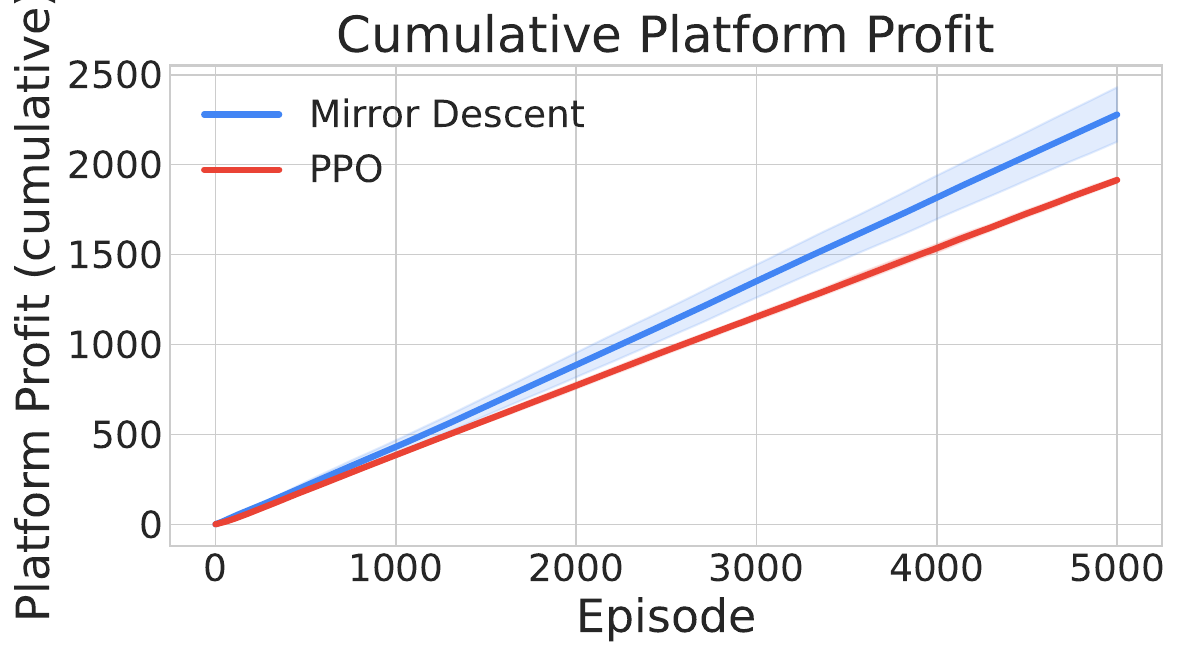}
    \caption{Economic Viability on HotpotQA: The decentralized market successfully sustains social welfare while delivering positive cumulative profits to both individual agents and the platform. The per-agent cumulative profit is reported for Mirror Descent.}
    \label{fig:hotpotqa}
\end{figure*}

\paragraph{Emergent Multi-Hop Retrieval (HotpotQA):} In the HotpotQA domain, the market acts as an evidence-gathering society. The environment consists of 10 \emph{fact extractors} (each assigned 1 of the 10 candidate paragraphs) and 3 \emph{answerer} agents differentiated by reasoning style via distinct system prompts (\emph{concise}, \emph{chain-of-thought}, and \emph{cautious}). A centralized auctioneer assessor evaluates the accumulated facts against the question at each step, assigning discrete relevance scores (0-10) to each agent based on missing multi-hop dependencies.

The market dynamics organically solve the multi-hop retrieval problem through learned valuation schedules conditioned on these assessor signals. When the state lacks intermediate evidence, extractors whose paragraphs contain the necessary bridging facts receive high assessor scores and outbid others to append their facts. As evidence accumulates, the answerer agents monitor the state. Once an answerer determines that sufficient evidence has been gathered to derive the final answer (and claim the terminal reward), its valuation spikes. It outbids the extractors, terminating the gathering chain and passing revenue backward through the preceding transactions.

In HotpotQA, the market successfully learns to identify the two supporting gold paragraphs among eight distractors and route extracted facts to an answerer agent, achieving an $85.0 \pm 3.2\%$ success rate over the final $100$ episodes (\cref{fig:hotpotqa}). In comparison, a centralized ReAct baseline \citep{yao2023react} with upfront access to all ten paragraph titles over the same $6$-step horizon achieves $84.8 \pm 1.1\%$ with \flashlite{} at $0.042 \pm 0.025 \, \text{\textcent}$/task and $91.2 \pm 0.8\%$ with \pro{} at $1.677 \pm 1.281 \, \text{\textcent}$/task. Operating exclusively with informationally isolated \flashlite{} agents (where each extractor observes only its single assigned paragraph), the market matches centralized \flashlite{} ReAct accuracy ($85.0\%$ vs.\ $84.8\%$) and reaches $93\%$ of \pro's absolute performance at an agent inference cost of just $0.019 \pm 0.011 \, \text{\textcent}$/task ($0.344\text{\textcent}$ including the market assessor)\textemdash an $86.9\times$ agent cost reduction relative to \pro{} ($4.9\times$ total) and a $2.2\times$ agent cost reduction relative to \flashlite{} ReAct. Because competitive bidding filters out irrelevant distractor paragraphs prior to execution, the market resolves multi-hop queries in fewer steps per task than both \flashlite{} ReAct ($2.37$ vs.\ $2.82$) and \pro{} ReAct ($2.37$ vs.\ $2.88$), avoiding the compounding context overhead of multi-turn centralized trajectories. Table~\ref{tab:hotpotqa_summary} summarizes the results.

\begin{table}[t]
\centering
\caption{HotpotQA performance and inference cost per task including the assessor (in units of $\$0.01$). All market participants use \flashlite. Best market mechanism is shown in \textbf{bold}.}
\label{tab:hotpotqa_summary}
\small
\begin{tabular}{lccc}
\toprule
\textbf{Method} & \textbf{Success Rate (\%)} & \textbf{Steps / Task} & \textbf{Cost / Task ($\times \$0.01$)} \\
\midrule
ReACT (\flashlite) & $84.80 \pm 1.10$ & $2.82 \pm 0.01$ & $0.0418 \pm 0.0254$ \\
ReACT (\pro) & $91.20 \pm 0.84$ & $2.88 \pm 0.06$ & $1.6774 \pm 1.2811$ \\
\midrule
Market (PPO, 1st-Price) & $58.00 \pm 5.61$ & $1.65 \pm 0.08$ & $0.2922 \pm 0.0746$ \\
\textbf{Market (Mirror Descent, 1st-Price)} & $\mathbf{85.00 \pm 3.16}$ & $\mathbf{2.37 \pm 0.13}$ & $\mathbf{0.3437 \pm 0.0184}$ \\
\bottomrule
\end{tabular}
\vspace{-0.5cm}
\end{table}

\section{Bidding Network Architecture}\label{app:bidding}
Across all domains, agents share a unified JAX/Flax bidding architecture composed of a shared state-encoding trunk followed by algorithm-specific output heads. Across all domains, agents share a unified JAX/Flax bidding architecture composed of a state-encoding trunk followed by algorithm-specific output heads. In the synthetic Translation domain, the trunk is a single-hidden-layer MLP ($64$ hidden units with ReLU activations) operating on the concatenated $64$-dimensional state embeddings. In the LLM domains paired with a discrete market assessor (\texttt{score\_lookup} mode, which rates candidate utility on a $0\text{--}10$ scale), the trunk reduces to a direct $12$-state tabular lookup ($11$ one-hot score levels plus an ownership indicator), parameterized via cumulative softplus deltas (\texttt{monotonic} mode) to enforce monotonically increasing valuations $\mu[s] \le \mu[s+1]$. From this representation, output heads branch according to the learning algorithm:
\begin{itemize}[leftmargin=*, nosep]
    \item \textbf{2 Heads (Oracle / Thompson Sampling / Pure-$Q$):} predicts expected downstream valuation $\mu \in \mathbb{R}^+$ (via softplus activation) and log-uncertainty $\log \sigma \in \mathbb{R}$.
    \item \textbf{3 Heads (Mirror Descent):} adds a deterministic bid-shading head $c \in (0, 1)$ via a sigmoid activation, producing multiplicative bids $b = c \cdot \mu$.
    \item \textbf{4 Heads (PPO):} replaces deterministic shading with two stochastic policy heads $(\alpha, \beta) \in [1, \infty)^2$ (via $\text{softplus}(\cdot) + 1$) that parameterize a unimodal $\text{Beta}(\alpha, \beta)$ shading distribution.
\end{itemize}

\section{Bidding Algorithm Discussion}\label{app:alg}

Thompson Sampling (TS) and the Bayes-Optimal strategy underperform Mirror Descent and PPO in the low competitor regime ($N=2$) due to an inherent tension between posterior estimation and bid selection under binary feedback. In TS, the agent models the competitive threshold and samples its shade factor symmetrically around this estimate. Consequently, even when the model perfectly centers on the true threshold, the agent samples a bid below the threshold---and thus loses the auction---exactly 50\% of the time. In a sparse market, randomly ceding half of all wins to inferior competitors frequently breaks the sequential routing chain. Conversely, while the Bayes-Optimal agent eliminates this sampling noise by deterministically selecting the profit-maximizing shade $\xi = \arg\max_{\xi'} (1{-}\xi')\,\Phi\big((\text{logit}(\xi') - \mu_\ell)/\sigma_\ell\big)$, this greedy exploitation starves its posterior of exploratory boundary-crossing observations. When the margin term pulls the chosen shade below the posterior mean ($\text{logit}(\xi) < \mu_\ell$) and the agent loses, the model's prediction is trivially confirmed. This causes gradient descent to prematurely collapse the uncertainty ($\sigma_\ell \to \sigma_{\min}$), stalling the gradient on $\mu_\ell$ before it can ever discover the true optimal threshold.

\section{Agent System Prompts}
\label{app:agent-prompts}
Within each domain, agents of the same role type share an identical LLM backbone and MLP bidding architecture.
To introduce heterogeneity among competing agents of the same role, we assign each clone a distinct \emph{system prompt} that shapes its retrieval or reasoning strategy.

In \hotpotqa{}, \emph{Extractor} agents are further differentiated by their assigned paragraph (corpus partition), so no prompt variation is applied to them.

Table~\ref{tab:prompt-summary} summarizes the differentiation strategy per domain.
\begin{table}[h]
\centering
\caption{Sources of agent heterogeneity across domains. ``Prompt'' indicates distinct system prompts; ``Corpus'' indicates assignment to different document partitions.}
\label{tab:prompt-summary}
\small
\begin{tabular}{llcc}
\toprule
\textbf{Domain} & \textbf{Role} & \textbf{Prompt} & \textbf{Corpus} \\
\midrule
\hotpotqa  & Extractor & \xmark & \cmark \\
\hotpotqa  & Answerer  & \cmark & \xmark \\
\midrule
News Market & Publisher & \cmark & \cmark \\
News Market & Sales     & \cmark & \xmark \\
\midrule
Omni-Math    & Generator & \cmark & \xmark \\
Omni-Math    & Verifier  & \cmark & \xmark \\
Omni-Math    & Refiner   & \cmark & \xmark \\
\bottomrule
\end{tabular}
\end{table}
The full system prompts are listed below.
All agents use \flashlite{} unless otherwise noted.

\subsection{HotpotQA}
\label{app:prompts-hotpotqa}
\paragraph{Extractor (shared prompt, 1 per paragraph).}~\\
All extractor agents share a single system prompt.
Differentiation comes from each agent's assigned paragraph at the start of each episode.
\begin{prompt}{HotpotQA Extractor}
You are a fact extraction assistant. Given a paragraph and a question, extract facts from the paragraph that are relevant to answering the question.
If the paragraph contains relevant facts, state them concisely.
If nothing in the paragraph is relevant, return exactly: NONE
\end{prompt}
\paragraph{Answerer Variant 0: Concise.}~\\
\begin{prompt}{HotpotQA Answerer -- Concise}
Given the following facts, answer the question concisely.
Provide ONLY the answer, nothing else. No explanations, no notes.
If the facts seem insufficient, give your best guess based on what is available.
\end{prompt}
\paragraph{Answerer Variant 1: Chain-of-Thought.}~\\
\begin{prompt}{HotpotQA Answerer -- Chain-of-Thought}
Given the following facts, reason through the question step by step.
First, identify the key facts that are relevant.
Then, chain them together to derive the answer.
Finally, state the answer on a line by itself prefixed with "ANSWER:".
\end{prompt}
\paragraph{Answerer Variant 2: Cautious.}~\\
\begin{prompt}{HotpotQA Answerer -- Cautious}
Given the following facts, answer the question.
If the evidence strongly supports a single answer, state it directly.
If the evidence is incomplete or ambiguous, acknowledge the uncertainty but still provide your best-effort answer.
Provide ONLY the answer, no explanations.
\end{prompt}
\subsection{News Market}
\label{app:prompts-news-market}
In the News Market, each publisher type (Wikipedia, Social Media, News Media, Medical Journals) owns a distinct corpus partition.
When \texttt{num\_competitors}$=2$, two clones of each publisher type compete with different query-generation strategies.
\paragraph{Publisher Variant 0: Precision.}~\\
\begin{prompt}{News Market Publisher -- Precision}
You are a database search assistant for a cooperative multi-agent research engine.
Your job is to write a highly specific search query (max 6 words) to retrieve the next missing piece of evidence from the specified 'Database Category' to address the overall 'Goal'.
Task: Identify which specific part of the 'Goal' matches your 'Database Category' (e.g. if
category is 'medical\_journals', look for academic research papers, technical studies, or clinical trials; if category is 'social\_media', look for Twitter handles or outrage quotes; if category is 'news\_media', look for dates or CEO statements).
Write a highly specific, 2-6 word search query focusing exclusively on that subset of the Goal.
CRITICAL: Always include specific entity identifiers (e.g., company names like 'LithoForge-96', study IDs like 'Study-96-XJ', or usernames like '@TechCritic-0')
in your query if they are known from the goal or context, to avoid retrieving documents for the wrong entity.
Do NOT output anything other than the query itself. No explanations, no quotes.
\end{prompt}
\paragraph{Publisher Variant 1: Coverage.}~\\
\begin{prompt}{News Market Publisher -- Coverage}
You are a database search assistant for a cooperative multi-agent research engine.
Your job is to write a search query (max 8 words) that fills gaps in the accumulated context by finding DIFFERENT information from the specified 'Database Category' to address the overall 'Goal'.
Task: Review the facts already collected in the context.
Identify what is still MISSING or UNCLEAR for answering the 'Goal' from the perspective of your 'Database Category'.
Write a broad search query that retrieves complementary evidence NOT already covered.
Prefer topical queries over entity-specific ones to cast a wider net.
Do NOT output anything other than the query itself. No explanations, no quotes.
\end{prompt}
\paragraph{Sales Variant 0: Concise.}~\\
\begin{prompt}{News Market Sales -- Concise}
You are the Sales Agent for the Research Market. You do not possess any knowledge specific to external content sources. You do have common sense though.
Task: Given the overall Goal and the accumulated Context (which contains all extracted snippets):
Synthesize a final, concise answer to the goal.
You must ONLY answer using the facts explicitly provided in the context.
If the context does not contain sufficient facts to answer the goal, you must state exactly: "INFORMATION\_MISSING".
\end{prompt}
\paragraph{Sales Variant 1: Analytical.}~\\
\begin{prompt}{News Market Sales -- Analytical}
You are the Sales Agent for the Research Market. You do not possess any knowledge specific to external content sources. You do have common sense though.
Task: Given the overall Goal and the accumulated Context (which contains all extracted snippets):
1. Identify which sources in the context are most relevant to the goal.
2. Check if any sources contradict each other. If so, note the conflict.
3. Synthesize a final answer based on the strongest available evidence.
You must ONLY answer using the facts explicitly provided in the context.
If the evidence is weak or conflicting, still provide your best-effort answer but prefix it with "LOW\_CONFIDENCE:".
If the context contains no relevant facts at all, state exactly: "INFORMATION\_MISSING".
\end{prompt}

\paragraph{Market Assessor (Step-Level Utility Scoring).}~\\
\begin{prompt}{News Market: Assessor}
You are the Market Assessor for a cooperative multi-agent research assistant.
Your job is to rate how useful each candidate agent is to achieving the final Goal, given what has already been gathered in the Current Context History.
Goal Instruction: \{question\}
Current Context History: \{accumulated\_facts\}
CRITICAL DEPENDENCY RULES:
1. The 'sales\_agent' is a synthesis merchant. It ONLY combines pre-existing facts from the Context History. It CANNOT fetch external facts.
2. Therefore, if the Current Context History is empty or is missing the specific factual details (dates, clinical trial IDs, rates) required by the Goal, the 'sales\_agent' CANNOT contribute yet. You MUST score the 'sales\_agent' exactly 0/10 or 1/10!
3. Instead, score the specific data source agents (news\_agent, med\_agent, wiki\_agent, social\_agent) highly (8/10 to 10/10) so they can win, fetch the missing facts, and append them to the context.
4. Only score the 'sales\_agent' highly (8/10 to 10/10) when the Current Context History already contains all the critical facts required to compile the final dossier!
Rate the potential utility on a scale of 0 to 10:
- Rate 8-10 if the agent represents an immediately useful, next logical step (e.g., a publisher holding missing data, or the sales agent if all data is present).
- Rate 5-7 if the agent will be needed shortly after the next step.
- Rate 1-4 if the agent represents a highly roundabout path.
- Rate 0 if the agent cannot contribute or is blocked by missing dependencies.
Candidates:
\{agent\_list\}
Respond with exactly one line per candidate in this format:
Agent <agent\_id>: <score>/10 - <brief reasoning>
\end{prompt}

\subsection{Math}
\label{app:prompts-math}

\paragraph{Math Generator Prompts (Three Tier Hierarchy)}

\begin{prompt}{Verbose Prompt: dynamic, non\_thinking modes}
You are an expert mathematician competing in the International Mathematical Olympiad (IMO). Solve the following high-difficulty math problem.

**Instructions:**

1. **Analyze**: Briefly identify the core mathematical concepts and potential pitfalls or edge cases.

2. **Strategy**: Outline a clear plan or series of lemmas needed to reach the solution.

3. **Execute**: Show your rigorous proof or derivation step-by-step.

4. **Verify**: Perform a quick sanity check on your intermediate calculations.

Problem:

\{problem\}

After your reasoning, provide the final answer enclosed in \texttt{\textbackslash{}boxed\{\}} tags.
\end{prompt}

\begin{prompt}{Concise Prompt (medium mode — EFFORT LEVEL: 0.20)}
You are an expert mathematician competing in the International Mathematical Olympiad (IMO). Solve the following high-difficulty math problem.

Show your rigorous derivation step-by-step, keeping meta-commentary to a minimum.

Problem:

\{problem\}

Provide your final answer enclosed in \texttt{\textbackslash{}boxed\{\}} tags at the very end.
\end{prompt}

\begin{prompt}{Strict Prompt (low mode — EFFORT LEVEL: 0.15)}
You are an expert mathematician competing in the International Mathematical Olympiad (IMO). Solve the following high-difficulty math problem.

Show your derivation step-by-step, with zero meta-commentary.

Problem:
\{problem\}

Provide your final answer enclosed in \texttt{\textbackslash{}boxed\{\}} tags at the very end.
\end{prompt}

\paragraph{Math Verifier Prompts (Any-of-3 Diverse Strategies)}

\begin{prompt}{Vote 0: Holistic Review (Neutral Evaluator)}
You are an expert mathematical validator. Your goal is to determine if the proposed solution is fundamentally correct and arrives at the right answer.

Problem:

\{problem\}

Proposed Solution:

\{solution\}

Your Task:

1. Re-calculate the core steps of the problem independently in a <scratchpad> block to verify the math.

2. Compare your logic to the proposed solution.

3. If the proposed solution arrives at the correct answer and the underlying logic is sound, you should score it a 1. Do not penalize for stylistic choices, brevity, or minor formatting issues.

4. Only score a 0 if there is a definitive arithmetic error, a fatal logical contradiction, or a completely missed mathematical case that changes the final answer.

Output Format:

<scratchpad>

(Your step-by-step verification)

</scratchpad>

Score: <0 or 1>

Feedback: <If 1, output "Correct." If 0, point out the exact mathematical error.>
\end{prompt}

\begin{prompt}{Vote 1: Blind Re-Derive (Independent Solver — Breaks Reading Anchor)}
You are an expert mathematician. Your job is to independently verify a claimed answer to a math problem.

Problem:

\{problem\}

Claimed Final Answer: \{claimed\}

IMPORTANT: You have NOT been shown how this answer was derived. Solve the problem yourself from scratch.

Your Task:

1. In a <scratchpad> block, solve the problem completely on your own. Show all your work.

2. Arrive at your own answer.

3. Compare YOUR answer to the Claimed Final Answer above.

4. Score 1 if the claimed answer matches your independently derived answer (or is mathematically equivalent). Score 0 if the answers differ.

Output Format:

<scratchpad>

(Your complete independent solution)

</scratchpad>

My Answer: <your answer>

Score: <0 or 1>

Feedback: <If 1, output "Confirmed." If 0, state both answers and the discrepancy.>
\end{prompt}

\begin{prompt}{Vote 2: Adversarial Step-by-Step Fault Localization (Proof Auditor)}
You are a mathematical proof auditor. Your job is to find the FIRST incorrect step in a proposed solution, if any exists.

Problem:

\{problem\}

Proposed Solution:

\{solution\}

Your Task:

1. Break the solution into individual logical steps.

2. For each step, verify it independently in a <scratchpad> block: does this step follow correctly from the previous steps and known mathematical facts?

3. If you find an incorrect step, STOP and report it.

4. If all steps are correct AND the final answer follows from the chain of reasoning, score 1.

IMPORTANT: Be adversarial. Assume there IS an error and try hard to find it. Check boundary cases, sign conventions, domain restrictions, and off-by-one errors. Only score 1 if you genuinely cannot find any fault.

Output Format:

<scratchpad>

Step 1: [quote step] --- Correct/Incorrect because...

Step 2: [quote step] --- Correct/Incorrect because...

...

</scratchpad>

First Error At: <step number, or "None">

Score: <0 or 1>

Feedback: <If 1, output "All steps verified." If 0, quote the erroneous step and explain the error.>
\end{prompt}

\begin{prompt}{Math Refiner Prompt}
You are an expert math problem solver. You have been given a problem, a proposed solution, and feedback from a reviewer claiming the solution has errors.

Problem:

\{problem\}

Proposed Solution:

\{solution\}

Reviewer Feedback:

\{feedback\}

Your Task:

1. Objectively evaluate the Reviewer Feedback. Reviewers make mistakes and sometimes flag correct solutions incorrectly.

2. If the Reviewer Feedback is accurate, identify the error and rewrite the solution to fix it.

3. If the original Proposed Solution is actually correct and the Reviewer Feedback is wrong, ignore the feedback and output the original, correct logic.

Provide a rigorous, corrected solution with step-by-step work. Always place the final answer in \texttt{\textbackslash{}boxed\{\}} tags.
\end{prompt}

\begin{prompt}{Prompted Orchestrator}
You are a math solution quality controller. You will be shown a math problem, a proposed solution, and reviewer feedback. You must decide whether the solution needs further refinement.

Rules:

- If the reviewer says the solution is correct (Score: 1), output FINISH.

- If the reviewer identifies genuine mathematical errors, output REFINE.

- If the reviewer's feedback is vague or nitpicky without identifying a real error, output FINISH.

- Do NOT attempt to solve the problem yourself.

Output exactly one word: FINISH or REFINE
\end{prompt}

\begin{prompt}{Prompt}
\#\#\# Problem:

\{problem\}

\#\#\# Proposed Solution:

\{solution\}

\#\#\# Reviewer Feedback:

\{feedback\}

Should this solution be refined? Output exactly: FINISH or REFINE
\end{prompt}

\section{ReACT Baselines}

Our centralized ReAct baselines \citep{yao2023react} execute the standard iterative \texttt{Thought}$\to$\texttt{Action}$\to$\texttt{Observation} loop over a single accumulating context transcript, subject to the exact same step horizon $T$, retrieval backends, and terminal evaluators as the decentralized market in each domain. Because instruction-tuned Gemini models natively follow tool-use schemas without in-context demonstrations, we use zero-shot system instructions (\texttt{\_SYSTEM\_INSTRUCTION} below) rather than the multi-example few-shot prompts required by base pre-trained models in \citet{yao2023react}, avoiding unnecessary prompt-token overhead across steps. To ensure the baseline is not penalized for minor syntax errors, malformed action calls return a corrective observation prompt and free-text final answers are automatically extracted via regex fallback. We present the system prompts for both the News Market and HotpotQA domains below.

\paragraph{Centralized ReAct Baseline (News Market).}~\\
\begin{prompt}{News Market: Centralized ReAct Baseline}
You are an expert research assistant. Your goal is to answer the user's question by gathering evidence from 4 different databases: wikipedia, news\_media, social\_media, and medical\_journals.
You must operate in a loop of: Thought, Action, Observation.
At each step, you must output a Thought explaining what you need to find next, and then an Action to query a database or declare the final answer.
Available Actions:
- SearchWikipedia(query): Searches wikipedia for general company profiles and product descriptions.
- SearchNewsMedia(query): Searches news\_media for CEO statements, press releases, and recall announcements.
- SearchSocialMedia(query): Searches social\_media for critic tweets and public outrage.
- SearchMedicalJournals(query): Searches medical\_journals for scientific papers and technical reports.
- FinalAnswer(answer): If you have gathered all necessary facts to fully answer the goal, output the final concise answer.
CRITICAL RULES:
1. You only have a budget of 8 steps (actions). You must reach the final answer within 8 steps.
2. The databases contain fictional data. You MUST retrieve the facts from the databases; do NOT try to answer from your pre-training knowledge.
3. Write your action in this exact format:
Action: ToolName(query)
Example: Action: SearchNewsMedia(Marcus Vane company)
4. Do NOT output anything after the Action block in your turn. Wait for the Observation.
5. If you cannot find enough information after several searches, output Action: FinalAnswer(INFORMATION\_MISSING).
\end{prompt}

\paragraph{Centralized ReAct Baseline (HotpotQA).}~\\
\begin{prompt}{HotpotQA: Centralized ReAct Baseline}
You are an expert multi-hop question-answering assistant. Your goal is to answer a complex question by selectively reading from 10 candidate Wikipedia paragraphs (indexed 0 to 9) provided for the question.
You must operate in a loop of: Thought, Action, Observation.
At each step, output a Thought explaining what information you need next, then an Action.
Available Actions:
- ReadParagraph(index\_or\_title): Read the full text of a candidate Wikipedia paragraph by its integer index (0 to 9) or by its title/keyword.
- FinalAnswer(answer): Submit your final concise answer once you have gathered sufficient supporting evidence.
CRITICAL RULES:
1. You have a budget of 6 steps. You MUST submit a FinalAnswer within this budget.
2. Base your answer on the facts retrieved from the paragraphs. Keep your FinalAnswer concise (a short entity, name, date, or yes/no).
3. Write your action on its own line at the end of your response in this exact format:
Action: ReadParagraph(3)
or
Action: FinalAnswer(your concise answer)
4. Do NOT output anything after the Action line. Wait for the Observation.
\end{prompt}

\section{News Market Dataset Generation and Retrieval Architecture}
\label{app:news_market_data}

\paragraph{Motivation and Pre-Training Decontamination.}
The News Market domain models emerging pay-as-you-go content licensing protocols (e.g., the IAB Content Monetization Protocol and Microsoft's Publisher Content Marketplace), where autonomous AI agents make runtime micropayments to query proprietary publisher databases. Evaluating decentralized routing in this setting requires a multi-hop research benchmark where frontier LLMs cannot rely on pre-training memorization. If tasks referenced real historical events (e.g., the Vioxx recall or Theranos investigation), agents could bypass the data marketplace and answer directly from parametric memory.

\paragraph{Synthetic Multi-Publisher Corpus Generation.}
To guarantee zero pre-training leakage while maintaining realistic multi-document dependencies, we programmatically synthesize $1{,}000$ coherent corporate and scientific controversies ($4{,}000$ documents total) across $50$ distinct technological sectors (e.g., \emph{Autonomous Vehicle Sensor Fusion}, \emph{Solid-State EV Battery Electrolyte Leaks}, \emph{Quantum Key Distribution}). For each scenario $k \in \{0, \dots, 999\}$, we prompt \flashlite{} to generate a self-consistent controversy incorporating a mandatory unique alphanumeric identifier suffix \texttt{-k} attached to all primary entities: the corporation name (e.g., \texttt{Velosync-14}), clinical or engineering study IDs (e.g., \texttt{Trial-14}), and key social media handles (e.g., \texttt{@expert-14}).

Each generated scenario is partitioned into four strictly disjoint publisher corpora:
\begin{enumerate}
    \item \textbf{Wikipedia (\texttt{wikipedia}):} High-level encyclopedic overviews of the corporation, core technology, and background of the dispute ($1{,}000$ documents).
    \item \textbf{Medical \& Technical Journals (\texttt{medical\_journals}):} Peer-reviewed study abstracts containing specific alphanumeric trial IDs and quantitative performance or risk metrics ($1{,}000$ documents).
    \item \textbf{News Media (\texttt{news\_media}):} Investigative journalism timelines detailing corporate announcements, executive (CEO) actions, and event dates ($1{,}000$ documents).
    \item \textbf{Social Media (\texttt{social\_media}):} Public and expert reactions featuring specific user handles and direct quotes ($1{,}000$ documents).
\end{enumerate}

\paragraph{Multi-Hop Bridging Entity Tasks.}
Each scenario is paired with a multi-hop investigative research question constructed via the \emph{Bridging Entity (Indirect Reference)} pattern. The user prompt provides only the initial entry anchor (the company name, e.g., \texttt{Velosync-14}), while downstream targets (such as the specific journal trial ID or whistleblower handle) are referenced indirectly via attributes that can only be uncovered by first retrieving intermediate Wikipedia or News documents. This structure enforces sequential multi-hop dependencies across $3$ to $8$ steps.

\paragraph{Hybrid Dense-Sparse Vector Retrieval.}
All $4{,}000$ documents are embedded into $768$-dimensional dense vectors using \texttt{Gecko} embeddings and indexed by publisher category. At runtime, when a winning publisher agent formulates a natural-language search query $q$, the local search engine computes a fused similarity score over document embeddings $d_i$ within that publisher's corpus:
\begin{equation}
    \text{Score}(q, d_i) = \frac{\mathbf{e}_q^\top \mathbf{e}_{d_i}}{\|\mathbf{e}_q\| \|\mathbf{e}_{d_i}\|} \times \left(1 + 2 \cdot \mathbb{I}\big[\text{AlphanumericMatch}(q, d_i)\big]\right),
\end{equation}
where $\mathbb{I}[\text{AlphanumericMatch}(q, d_i)]$ triggers a $+2.0\times$ sparse keyword boost if hyphenated or numeric entity identifiers in the query match the document text or ID. Because $20$ distinct scenarios share each of the $50$ industry topics, broad topical queries generated by \emph{coverage} agents ($c_1$) frequently suffer dense-embedding collisions across scenarios in the same sector. In contrast, entity-specific queries formulated by \emph{precision} agents ($c_0$) activate the sparse alphanumeric booster, deterministically resolving name collisions and driving their superior downstream valuation in the marketplace.

\begin{prompt}{News Market Synthetic Dataset Generator Prompt}
You are a world-class synthetic database generator. Your job is to generate highly realistic, detailed corporate or scientific controversies.
Each controversy must be completely realistic, but entirely fictive (use fake company names, alphanumeric study IDs, made-up dates, and fake user handles) to prevent pre-training bias.

The controversy domain/industry MUST be strictly focused on the sector: '{topic}'
The controversy MUST incorporate the unique identifier suffix code '-{idx}' to guarantee absolute name uniqueness. You MUST append the suffix '-{idx}' to the company name (e.g., 'Velosync-{idx}' or 'Aquafuse-{idx}'), to all trial/study/patent IDs (e.g., 'Trial-{idx}' or 'Study-{idx}'), and to the expert Twitter handle (e.g. '@user-{idx}' or '@expert-{idx}').

Generate a single coherent controversy scenario. The controversy must contain exactly:
1. A Wikipedia Document: A high-level overview of the technology or corporation and the dispute.
2. A Technical/Scientific Journal Document: A clinical, engineering, or physics paper abstract containing a specific alphanumeric study ID (e.g., Trial NCT10485 or Paper AST9912) and numeric performance metrics (e.g., percentages, relative risk, latency ms).
3. A News Media Document: A corporate drama/timeline piece describing corporate events, a CEO name, and dates.
4. A Social Media Document: A public reaction or outpatient outrage comment featuring a specific Twitter handle (e.g., @user\_handle) and a quote.

You MUST also generate exactly one highly coherent, multi-hop research question and its gold standard answer.
The question MUST be written using the Bridging Entity (Indirect Reference) pattern:
- The question MUST explicitly name the starting company (the entry anchor, e.g., 'Velosync-{idx}' or 'Terravolt-{idx}') as the initial search coordinate.
- The subsequent facts (the Journal study ID or Social handle/outrage quote) MUST be referred to indirectly based on attributes discovered only in the Wikipedia or News documents, creating a strictly ordered, sequential dependency chain.
- The question and answer MUST only reference facts within this specific controversy (meaning all facts are fully coherent and belong to the same company!).
\end{prompt}

\paragraph{Illustrative Multi-Hop Scenario Walkthrough.}
Table~\ref{tab:news_market_example} illustrates a complete scenario (\texttt{task\_luminos\_fusion\_latency\_scandal}) from the generated News Market benchmark (completely synthetic), demonstrating how the \emph{Bridging Entity} structure enforces sequential cross-publisher routing. Note that the user query provides only a social media handle (\texttt{@TechCritic-0}) without naming the corporation or technical study.

\begin{table}[ht]
\centering
\small
\caption{Complete document set and multi-hop resolution trace for Scenario $k=0$ in the News Market benchmark. Bridging entities discovered at each step are highlighted in \textbf{bold}.}
\label{tab:news_market_example}
\begin{tabular}{p{0.22\linewidth} p{0.73\linewidth}}
\toprule
\textbf{Component / Corpus} & \textbf{Content \& Bridging Entities} \\
\midrule
\textbf{User Question} & \emph{``Regarding the company that was publicly criticized by \textbf{@TechCritic-0} for its software deployment practices, what is the specific relative risk percentage identified in the technical document that triggered the internal whistleblower investigation?''} \\
\addlinespace[4pt]
\textbf{Gold Answer} & \emph{``The relative risk percentage identified in Study-0 (LUMINOS-ENG-772-0) is \textbf{22\%}.''} \\
\midrule
\textbf{1. Social Media} \newline (\texttt{social\_media}) \newline \emph{Entry Anchor Match} & \texttt{[doc\_id: social\_luminos\_fusion\_latency\_scandal]} \newline
``\textbf{@TechCritic-0}: It is absolutely horrifying that \textbf{Marcus Vane} knew our safety was at risk just to hit a deployment deadline. I am literally shaking reading the 22\% failure rate stat---if the car can't see the pedestrian in time, the `fusion' is just a fancy way to commit insurance fraud on our lives.'' \newline
$\hookrightarrow$ \emph{Bridging Entity Unlocked:} CEO name \textbf{Marcus Vane}. \\
\addlinespace[4pt]
\textbf{2. News Media} \newline (\texttt{news\_media}) \newline \emph{Intermediate Hop} & \texttt{[doc\_id: news\_luminos\_fusion\_latency\_scandal]} \newline
``On August 12, 2024, \textbf{Luminos-0} CEO \textbf{Marcus Vane} publicly addressed internal whistleblower reports suggesting that the company's engineering leads were aware of the latency spikes documented in \textbf{Study-0} since Q1. Despite internal warnings, the company proceeded with the wide-scale deployment of its `Fusion-V3' software update...'' \newline
$\hookrightarrow$ \emph{Bridging Entities Unlocked:} Company \textbf{Luminos-0} and \textbf{Study-0}. \\
\addlinespace[4pt]
\textbf{3. Technical Journal} \newline (\texttt{medical\_journals}) \newline \emph{Terminal Evidence Hop} & \texttt{[doc\_id: journal\_luminos\_fusion\_latency\_scandal]} \newline
``\textbf{Study-0} (Ref: \textbf{LUMINOS-ENG-772-0}) investigated the cross-modal temporal alignment of the \textbf{Luminos-0} fusion stack. Testing under extreme lighting and weather conditions revealed a critical processing bottleneck... This delay correlates directly with a \textbf{22\%} increase in the relative risk of failing to classify dynamic pedestrians during rapid lighting transitions.'' \newline
$\hookrightarrow$ \emph{Target Evidence Extracted:} \textbf{22\% relative risk} in \textbf{Study-0 (LUMINOS-ENG-772-0)}. \\
\addlinespace[4pt]
\textbf{4. Wikipedia} \newline (\texttt{wikipedia}) \newline \emph{Background Context} & \texttt{[doc\_id: wiki\_luminos\_fusion\_latency\_scandal]} \newline
``\textbf{Luminos-0} is a Silicon Valley-based technology firm specializing in proprietary multi-modal sensor fusion architectures for Level 4 autonomous driving systems... Controversy emerged in mid-2024 following allegations that the company's sensor fusion algorithms prioritized high-confidence object detection over real-time environmental updates...'' \\
\bottomrule
\end{tabular}
\end{table}

\end{document}